\documentclass[10pt,a4paper]{article}
\usepackage{amssymb}
\usepackage{amsmath}
\usepackage{braket}
\usepackage[ruled,noend]{algorithm2e}
\usepackage{float}

\usepackage{enumitem}
\usepackage{multirow}

\usepackage{tabularx}
\usepackage{rotating}
\newcounter{subroutine}
\newenvironment{subroutine}[1]
{\par\addvspace{\topsep}
	\noindent
	\tabularx{\linewidth}{@{} X @{}}
	\hline
	\refstepcounter{subroutine}\textbf{Subroutine \thesubroutine} #1 \\
	\hline}
{ \\
	\hline
	\endtabularx
	\par\addvspace{\topsep}}

\newcommand{\sbline}{\\[.5\normalbaselineskip]}

\usepackage{capt-of}
\providecommand{\keywords}[1]
{
	\small    
	\textbf{\textit{Keywords--}} #1
}

\begin{document}
\setlength{\textheight}{8.0truein}    


\thispagestyle{empty}
\setcounter{page}{1}


\vspace*{0.88truein}



\centerline{\bf
SECURE MULTI-PARTY QUANTUM CONFERENCE AND XOR COMPUTATION}
\vspace*{0.37truein}
\centerline{\footnotesize
NAYANA DAS \footnote{E-mail: dasnayana92@gmail.com}}
\vspace*{0.015truein}
\centerline{\footnotesize\it Applied Statistics Unit, Indian Statistical Institute}
\baselineskip=10pt
\centerline{\footnotesize\it Kolkata 700108, India}
\vspace*{10pt}
\centerline{\footnotesize 
GOUTAM PAUL\footnote{E-mail: goutam.paul@isical.ac.in}}
\vspace*{0.015truein}
\centerline{\footnotesize\it Cryptology and Security Research Unit, R. C. Bose Centre for Cryptology and Security}
\baselineskip=10pt
\centerline{\footnotesize\it Indian Statistical Institute, Kolkata 700108, India}
\vspace*{0.225truein}

\vspace*{0.21truein}

\begin{abstract}
		Quantum conference is a process of securely exchanging messages between three or more parties, using quantum resources. A Measurement Device Independent Quantum Dialogue (MDI-QD) protocol, which is secure against information leakage, has been proposed (Quantum Information Processing 16.12 (2017): 305) in 2017, is proven to be insecure against intercept-and-resend attack strategy. We first modify this protocol and generalize this MDI-QD to a three-party quantum conference and then to a multi-party quantum conference. We also propose a protocol for quantum multi-party XOR computation. None of these three protocols proposed here use entanglement as a resource and we prove the correctness and security of our proposed protocols.
\end{abstract}

\vspace*{3pt}
\section{Introduction}
\label{intro}
In the current post-digital era, quantum cryptography has generated significant interest in the information security domain. Security of quantum cryptographic protocols mainly depends on the ``no-cloning theorem"~\cite{wootters1982single} and the fact that, without disturbance, two non-orthogonal states can not be distinguished with a finite number of samples. The first-ever quantum cryptographic protocol was BB84 quantum-key-distribution (QKD), proposed by Bennett and Brassard in 1984~\cite{tcs/BennettB14}. QKD allows two or more remote users to establish a shared secret key between themselves. In BB84 protocol, two users, namely, Alice and Bob, exchange single-qubit states to generate a secret key. In 2000, Shor and Preskill showed that this protocol is secure and they gave a simple proof of security of the BB84 protocol~\cite{shor2000simple}. In 1991, Ekert proposed another QKD protocol using entangled states~\cite{ekert1991quantum}. Till now, there are many variants of QKD protocols proposed by many researchers, for example, BBM92~\cite{Brassard1992quantum}, B92~\cite{bennett1992quantum} and many others~\cite{long2002theoretically,xue2002conditional,deng2004bidirectional,hwang2003quantum,lo2005decoy,lo2012measurement,barrett2005no,grosshans2003quantum}.

Quantum secure direct communication (QSDC) is another nice part of quantum cryptography, whose purpose is to securely send a secret message from one party (Alice) to another party (Bob), without using any shared key. The famous ping-pong-protocol~\cite{bostrom2002ping} is an example of QSDC protocol, where the receiver Bob prepares two-qubit entangled states and sends one qubit to the sender Alice. Then Alice performs some unitary operations on that qubit to encode her information and sends it back to Bob. By measuring the joint state, Bob gets the message. Recently, other QSDC protocols with different approaches are also explored~\cite{deng2003two,deng2004secure,wang2005quantum,wang2005multi,wang2006quantum,long2007quantum,xi2007quantum,das2020improving,das2020cryptanalysis}.

A two-way QSDC protocol is called quantum dialogue (QD), where Alice and Bob can simultaneously exchange their messages with a single channel, was proposed by BA Nguyen in 2004~\cite{nguyen2004quantum}. Since then, many QD protocols ware proposed~\cite{zhang2004deterministic,zhong2005quantum,xia2006quantum,xin2006secure,yan2007controlled,tan2008classical,gao2008revisiting,gao2010two,qip/Maitra17,das2020two}. In~\cite{qip/Maitra17}, authors proposed a measurement device independent QD (MDI-QD) with the help of an untrusted third party (UTP) and showed that this protocol is secure against information leakage.

QSDC protocols for more than two parties are discussed in~\cite{gao2005deterministic,jin2006three,ting2005simultaneous,tan2014multi,zhang2005multiparty,banerjee2018quantum}. In~\cite{banerjee2018quantum}, the authors proposed the concept of quantum conference or $N$-party communication, $N\geq 2$, where each party sends their message to the other $(N-1)$ parties. In this protocol, to communicate $m$-bit classical messages, they need at least $(N-1)$ pairwise disjoint subgroups of unitary operators, where the cardinality of each subgroup is at-least $2^m$. For large $m$, finding these subgroups is quite difficult. 

All the above primitives are multi-party protocols, but not multi-party computation. In the multi-party protocol, two or more parties exchange messages over a public channel and perform some local computation to achieve a communication task. On the other hand, in multi-party computation, two or more parties exchange messages over a public channel and perform some local computation to jointly compute the value of a function on their private data as inputs. The requirement is that, after the end of the computation, each party will have the output of the function, but no party will have access to the input of any other party. Quantum multi-party computation (QMPC) is an interesting research area in quantum cryptography, where the parties possess some quantum states as inputs. Quantum secret sharing (QSS)~\cite{hillery1999quantum,zhang2005multiparty,zhang2005multiparty_qss,gottesman2000theory,guo2003quantum}, QMPC protocol for summation and multiplication~\cite{shi2016secure,chen2010}, quantum private comparison~\cite{Liu2013,Zhang2013,liu2015} are some examples of QMPC protocols.

\subsection*{Our Contribution}
In this paper, we make four distinct contributions. First, we revisit the two-party MDI-QD protocol~\cite{qip/Maitra17}, and show that this is not secure against intercept-and-resend attack. Then we modify the two-party MDI-QD protocol to make it secure against this attack. Second, using a similar approach, we propose a three-party quantum conference protocol with the help of an untrusted fourth party. Next, we generalize our three-party quantum conference protocol to a multi-party version. We show that both these conference protocols are correct and secure against  intercept-and-resend attack, entangle-and-measure attack, Denial-of-Service (DoS) attack and man-in-the-middle attack.  As the fourth and final contribution, we show how to use part of our multi-party quantum conference protocol to compute multi-party XOR function, and establish it's correctness and security.

\subsection*{Outline}
The rest of this paper is organized as follows: in Section~\ref{sec2}, we revisit the MDI-QD protocol proposed in~\cite{qip/Maitra17}. Then in the next section, we discuss intercept-and-resend attack on the MDI-QD protocol~\cite{qip/Maitra17} and we propose its modified version. Section~\ref{sec3} describes our proposed protocol for a three-party quantum conference and it's correctness and security analysis. We generalize our three-party quantum conference to $N$-party quantum conference in Section~\ref{sec4}. Next, we present a protocol for multi-party XOR computation, by using tools of $N$-party quantum conference in Section~\ref{sec5}. Section~\ref{sec6} concludes our results.

\subsection*{Notations}
Here we describe the common notations that will be used throughout the paper.
\begin{itemize}[label=$\bullet$]
\item $\ket{+}=\frac{1}{\sqrt{2}}(\ket{0}+ \ket{1})$, $\ket{-}=\frac{1}{\sqrt{2}}(\ket{0}- \ket{1})$;
\item $Z$ basis $=\{\ket{0},\ket{1}\}$;
\item $X$ basis $=\{\ket{+},\ket{-}\}$;
\item $\{S[i]\}_{i=1}^{m}=S$ is a finite sequence of length $m$;
\item $S[i]=S_i=i$-th element of $S$ ;
\item $\bar{b}$= bit complement of $b$;
\item $i_1i_2\ldots i_N = N$ bit binary representation of $i$;
\item $\ket{i}=\ket{i_1}\ket{i_2}\ldots \ket{i_N}$ is an $N$-qubit state;
\item $\ket{\Phi^{+}}=\frac{1}{\sqrt{2}}(\ket{00}+ \ket{11})$, $\ket{\Phi^{-}}=\frac{1}{\sqrt{2}}(\ket{00}- \ket{11})$;
\item $\ket{\Psi^{+}}=\frac{1}{\sqrt{2}}(\ket{01}+ \ket{10})$, $\ket{\Psi^{-}}=\frac{1}{\sqrt{2}}(\ket{01}- \ket{10})$;
\item $\mathcal{B}_N=\{\ket{\Phi_{0}^{+}}, \ket{\Phi_{0}^{-}},\ket{\Phi_{1}^{+}},\ket{\Phi_{1}^{-}}, \ldots, \ket{\Phi_{2^{(N-1)}-1}^{+}}, \ket{\Phi_{2^{(N-1)}-1}^{-}}\}$ basis,\\ where  $\ket{\Phi_{i}^{\pm}}=\frac{1}{\sqrt{2}}(\ket{i}\pm \ket{2^N-1-i})$ for $i \in \{0,1,\ldots ,2^{(N-1)}-1\}$.\\ For example : 
\begin{enumerate}
	\item $\mathcal{B}_2=\{\ket{\Phi_{0}^{+}}, \ket{\Phi_{0}^{-}},\ket{\Phi_{1}^{+}},\ket{\Phi_{1}^{-}}\}$ is called Bell basis; where 
	\begin{itemize}
		\item $\ket{\Phi_{0}^{+}}=\frac{1}{\sqrt{2}}(\ket{00}+ \ket{11})=\ket{\Phi^{+}}$, $\ket{\Phi_{0}^{-}}=\frac{1}{\sqrt{2}}(\ket{00}- \ket{11})=\ket{\Phi^{-}}$
		\item $\ket{\Phi_{1}^{+}}=\frac{1}{\sqrt{2}}(\ket{01}+ \ket{10})=\ket{\Psi^{+}}$, $\ket{\Phi_{1}^{-}}=\frac{1}{\sqrt{2}}(\ket{01}- \ket{10})=\ket{\Psi^{-}}$
	\end{itemize}
	
	\item $\mathcal{B}_3=\{\ket{\Phi_{0}^{+}}, \ket{\Phi_{0}^{-}},\ket{\Phi_{1}^{+}},\ket{\Phi_{1}^{-}}, \ket{\Phi_{2}^{+}}, \ket{\Phi_{2}^{-}}, \ket{\Phi_{3}^{+}}, \ket{\Phi_{3}^{-}}\}$ basis; where 
	\begin{itemize}
		\item $\ket{\Phi_{0}^{+}}=\frac{1}{\sqrt{2}}(\ket{000}+ \ket{111})$, $\ket{\Phi_{0}^{-}}=\frac{1}{\sqrt{2}}(\ket{000}- \ket{111})$
		\item $\ket{\Phi_{1}^{+}}=\frac{1}{\sqrt{2}}(\ket{001}+ \ket{110})$, $\ket{\Phi_{1}^{-}}=\frac{1}{\sqrt{2}}(\ket{001}- \ket{110})$
		\item $\ket{\Phi_{2}^{+}}=\frac{1}{\sqrt{2}}(\ket{010}+ \ket{101})$, $\ket{\Phi_{2}^{-}}=\frac{1}{\sqrt{2}}(\ket{010}- \ket{101})$
		\item $\ket{\Phi_{3}^{+}}=\frac{1}{\sqrt{2}}(\ket{011}+ \ket{100})$, $\ket{\Phi_{3}^{-}}=\frac{1}{\sqrt{2}}(\ket{011}- \ket{100})$;
	\end{itemize}
	
\end{enumerate}
\item $\Pr(A)=$ Probability of occurrence of an event $A$;
\item $\Pr(A|B)=$ Probability of occurrence of an event $A$ given that the event $B$ has already occurred;
\item $wt(v)=$ number of 1's in a binary vector $v$.

\end{itemize}

\section{Revisiting the Measurement Device Independent \\Quantum Dialogue (MDI-QD) Protocol of~\cite{qip/Maitra17} 
} \label{sec2}
Here, in this section, we shortly describe the MDI-QD protocol proposed in~\cite{qip/Maitra17}, where two legitimate parties, namely Alice and Bob, can simultaneously exchange their messages. The proposal in~\cite{qip/Maitra17} composed two different protocols from~\cite{tcs/BennettB14} and~\cite{lo2012measurement}. Alice and Bob first perform some QKD, namely, BB84~\cite{tcs/BennettB14} and generate a shared key $k$ between themselves. Then they prepare their sets of qubits ${Q_A}$ and $Q_B$, corresponding to $k$ and their respective messages $a$ and $b$. Alice and Bob send ${Q_A}$ and $Q_B$ to an untrusted third party or UTP (who may be an Eavesdropper). Then the UTP measures the two qubit states in Bell basis (i.e, $\mathcal{B}_2$) and announces the result. From the result, Alice and Bob decode the messages of each other (see Table~\ref{table_qd}). Details are given in Figure~\ref{algo:1qd}.
\begin{table}[h]
\centering
\caption{Different cases in MDI QD.}
\begin{tabular}{|c|c|c|c|c|c|c|c|}
	\hline
	\multicolumn{2}{|c|}{Bits to communicate by} &  \multicolumn{2}{|c|}{Qubits prepared by} & \multicolumn{4}{|c|}{Probabilities of measurement}\\
	\multicolumn{2}{|c|}{} &  \multicolumn{2}{|c|}{} & \multicolumn{4}{|c|}{results at UTP's end}\\
	\hline
	{} Alice  &  Bob   & Alice (${Q_A}_i$)  &  Bob (${Q_B}_i$) & $\ket{\Phi^+}$ & $\ket{\Phi^-}$ & $\ket{\Psi^+}$ & $\ket{\Psi^-}$ \\
	\hline
	$0$ & $0$  & $\ket{0}$  &  $\ket{0}$ & $1/2$ & $1/2$  & $0$ & $0$  \\
	$0$ & $1$ & $\ket{0}$  &  $\ket{1}$ &  $0$ & $0$ & $1/2$ & $1/2$  \\
	$1$ & $0$ & $\ket{1}$  &  $\ket{0}$ &  $0$ & $0$ & $1/2$ & $1/2$  \\
	$1$ & $1$ & $\ket{1}$  &  $\ket{1}$ & $1/2$ & $1/2$  & $0$ & $0$ \\
	\hline
	$0$ & $0$ & $\ket{+}$  &  $\ket{+}$ & $1/2$ & $0$  & $1/2$ & $0$ \\
	$0$ & $1$ & $\ket{+}$  &  $\ket{-}$ & $0$ & $1/2$  & $0$ & $1/2$ \\
	$1$ & $0$ & $\ket{-}$  &  $\ket{+}$ & $0$ & $1/2$  & $0$ & $1/2$ \\
	$1$ & $1$ & $\ket{-}$  &  $\ket{-}$ & $1/2$ & $0$  & $1/2$ & $0$\\
	\hline
\end{tabular}
\label{table_qd}
\end{table}

\begin{algorithm}[!tb]
\begin{enumerate}
	\item Alice and Bob share an $n$-bit key stream ($k=k_1k_2\ldots k_n$) between themselves using BB84 protocol.
	
	\item Let the $n$-bit message of Alice (Bob) be  $a=a_1a_2\ldots a_n$ ($b=b_1b_2\ldots b_n$).
	
	\item For $1\leq i \leq n$, Alice (Bob) prepares the qubits $Q_A={Q_A}_1{Q_A}_2\ldots {Q_A}_n ~(Q_B={Q_B}_1{Q_B}_2\ldots {Q_B}_n)$ at her  (his) end according to the following strategy:
	\begin{enumerate}
		\item if $a_i$ ($b_i $)$ = 0$ and $k_i = 0 \Rightarrow {Q_A}_i~({Q_B}_i)=\ket{0}$;
		\item if $a_i$ ($b_i $)$ = 1$ and $k_i = 0 \Rightarrow {Q_A}_i~({Q_B}_i)=\ket{1}$;
		\item if $a_i$ ($b_i $)$ = 0$ and $k_i = 1 \Rightarrow {Q_A}_i~({Q_B}_i)=\ket{+}$;
		\item if $a_i$ ($b_i $)$ = 1$ and $k_i =1 \Rightarrow {Q_A}_i~({Q_B}_i)=\ket{-}$.
	\end{enumerate}
	
	\item Alice (Bob) sends her (his) prepared qubits $Q_A~(Q_B)$ to an untrusted third party (UTP). 
	
	\item For $1\leq i \leq n$, the UTP measures each two qubits ${Q_A}_i$ and ${Q_B}_i$ in Bell basis (i.e., $\mathcal{B}_2=\{\ket{\Phi^+}\ket{\Phi^-},\ket{\Psi^+},\ket{\Psi^-}\}$) and announces the measurement result $\mathcal{M}_i \in \{\ket{\Phi^+}\ket{\Phi^-},\ket{\Psi^+},\ket{\Psi^-}\}$ publicly. Table~\ref{table_qd} shows the possible measurements results with their occurring probabilities. 
	
	\item For $1\leq i \leq n$, Alice and Bob consider the $i$-th measurement result $\mathcal{M}_i$, if $\mathcal{M}_i=\ket{\Phi^-}$ or $\ket{\Psi^+}$ and discard the other cases.  
	
	\item They randomly choose $\delta n$ number of measurement results to estimate the error,\\ where $\delta \ll 1$ is a small fraction.
	
	\item Alice and Bob guess the message bits of other, corresponding to their chosen $\delta n$ number of measurement results using Table~\ref{tab:Alice' guess about $b_i$ } and Table~\ref{tab: Bob's guess about $a_i$ }. 
	
	\item For the above mentioned $\delta n$ rounds, they disclose their respective guesses.
	
	\item If the estimated error is greater than some predefined threshold value, then they abort. Else they continue and go to the next step.
	
	\item For the remaining measurement results, Alice and Bob guess the message bits of \\each other, using Table~\ref{tab:Alice' guess about $b_i$ } and Table~\ref{tab: Bob's guess about $a_i$ }.
	
\end{enumerate}
\captionof{figure}{MDI-QD Protocol of~\cite{qip/Maitra17}}
\label{algo:1qd}
\setlength{\textfloatsep}{0.05cm}
\setlength{\floatsep}{0.05cm}
\end{algorithm}

	It is clear from Table~\ref{table_qd} that, for $1 \leq i \leq n$,
	\begin{itemize}
		\item if Alice prepares ${Q_A}_i=\ket{0}$($\ket{1})$, then she guesses $b_i$ with probability $1$ as follows:
		\begin{equation*}
		\mathcal{M}_i=
		\begin{cases}
		\ket{\Phi^+}$ or $\ket{\Phi^-} \Rightarrow &\text{$b_i=$ $0$ ($1$)};\\
		\ket{\Psi^+}$ or $\ket{\Psi^-} \Rightarrow &\text{$b_i=$ $1$ ($0$)},
		\end{cases}
		\end{equation*}
		
		\item if Alice prepares ${Q_A}_i=\ket{+}$($\ket{-})$, she guesses $b_i$ with probability $1$ as follows:
		\begin{equation*}
		\mathcal{M}_i=
		\begin{cases}
		\ket{\Phi^+}$ or $\ket{\Psi^+} \Rightarrow &\text{$b_i=$ $0$ ($1$)};\\
		\ket{\Phi^-}$ or $\ket{\Psi^-} \Rightarrow &\text{$b_i=$ $1$ ($0$)}.
		\end{cases}
		\end{equation*}
		
	\end{itemize}

	From the above discussion and Table~\ref{table_qd}, let us construct two more tables, namely Table~\ref{tab:Alice' guess about $b_i$ } and Table~\ref{tab: Bob's guess about $a_i$ }, containing the information of Alice's guess and Bob's guess about other's message bits for different cases.
	\begin{table}[h]
		\centering
		\caption{Alice's guess about Bob's message bit for different cases.}
		\begin{tabular}{|c|c|c|c|c|c|c|}
			\hline
			Key  & Alice's & Alice's &  \multicolumn{4}{|l|}{Alice's guess about $b_i$ when $\mathcal{M}_i$}\\
			\cline{4-7}
			bit $k_i$ & bit $a_i$ & qubit ${Q_A}_i$ & $\ket{\phi^+}$ &$\ket{\phi^-}$ &$\ket{\psi^+}$ &$\ket{\psi^-}$\\
			\hline
			0 & 0 & $\ket{0}$  & 0 & 0 & 1 & 1 \\
			
			0 & 1 & $\ket{1}$  & 1 & 1 & 0 & 0 \\
			
			1 & 0 & $\ket{+}$  & 0 & 1 & 0 & 1 \\
			
			1 & 1 & $\ket{-}$  & 1 & 0 & 1 & 0 \\
			\hline
		\end{tabular}

		\label{tab:Alice' guess about $b_i$ }
	\end{table}
	
	\begin{table}[h]
		\centering
		\caption{Bob's guess about Alice's message bit for different cases.}
		
		\begin{tabular}{|c|c|c|c|c|c|c|}
			\hline
			Key  & Bob's & Bob's &  \multicolumn{4}{|c|}{Bob's guess about $a_i$ when $\mathcal{M}_i$}\\ \cline{4-7}
			bit $k_i$ & bit $b_i$ & qubit ${Q_B}_i$ & $\ket{\phi^+}$ & $\ket{\phi^-}$ & $\ket{\psi^+}$ & $\ket{\psi^-}$\\
			\hline
			0 & 0 & $\ket{0}$  & 0 & 0 & 1 & 1\\
			
			0 & 1 & $\ket{1}$  & 1 & 1 & 0 & 0\\
			
			1 & 0 & $\ket{+}$  & 0 & 1 & 0 & 1\\
			
			1 & 1 & $\ket{-}$  & 1 & 0 & 1 & 0\\
			\hline
		\end{tabular}

		\label{tab: Bob's guess about $a_i$ }
	\end{table}

	Hence from Table~\ref{tab:Alice' guess about $b_i$ } and Table~\ref{tab: Bob's guess about $a_i$ }, we can say that both Alice and Bob can exchange their message simultaneously.
	
	Now, we can see from Table~\ref{table_qd}, for $1 \leq i \leq n$, if $\mathcal{M}_i= \ket{\Phi^+}$ or $\ket{\Psi^-}$, then Eve knows the information whether $a_i=b_i$ or not. That is, Eve knows $a_i \oplus b_i$ ($1$ bit of information out of 2 bits), for those $\mathcal{M}_i$, where $\mathcal{M}_i= \ket{\Phi^+}$ or $\ket{\Psi^-}$. To avoid this information leakage, Alice and Bob discard these cases. Then they estimate the error and if the error exceeds some predefined threshold, they abort the protocol. Otherwise, they continue it and guess other's message. 
	
	\section{Intercept-and-Resend Attack on the MDI-QD Protocol of~\cite{qip/Maitra17} and Our Proposed Remedy}
	We now show that the above MDI-QD protocol~\cite{qip/Maitra17} is not secure against intercept-and-resend attack and an adversary can get hold of some amount of information about the messages. So we propose a modified version of this protocol, which is secure against this attack. 
	
	Let us consider the intercept-and-resend attack by an adversary $\mathcal{A}$ (other than the UTP). For the $i$-th message bit pair $(a_i,b_i)$ of Alice and Bob, they prepare the qubit pair $(Q_{A_i},Q_{B_i})$ depending upon the key bit $k_i$, and send those qubits $Q_{A_i},Q_{B_i}$ to the UTP by separate channels from Alice and Bob. Now $\mathcal{A}$ intercepts the qubits $Q_{A_i},~Q_{B_i}$ from the channel and guesses the corresponding key bit ${k'}_i$ to choose the measurement basis for the qubits. $\mathcal{A}$ measures $Q_{A_i}$ and $Q_{B_i}$ in the same basis and resends those qubits to the UTP. Note that, if $\mathcal{A}$ guesses the correct key bit, then she chooses the correct basis to measure $Q_{A_i},~Q_{B_i}$, and due to this measurement, the states of the qubits remain unchanged. In this case, $\mathcal{A}$ gets the correct message bit-pair of Alice and Bob, without introducing any error in the channel. Now, if $\mathcal{A}$ chooses the wrong key bit, then also she can get the correct message bit-pair $(a_i,b_i)$ with probability $1/4$ and in this case, $\mathcal{A}$ can be detected with probability $1/2$. 
	
	As an illustrative example, consider $k_i=0$, ${k'}_i=1$, $a_i=0$, $b_i=0$, then $Q_{A_i}=\ket{0}$, $Q_{B_i}=\ket{0}$. Since ${k'}_i=1$, $\mathcal{A}$ measures $Q_{A_i},~Q_{B_i}$ in $X$-basis. After the measurement, let the qubits be ${Q'}_{A_i},~{Q'}_{B_i}$. If ${Q'}_{A_i}=\ket{+},~{Q'}_{B_i}=\ket{+}$, then also $\mathcal{A}$ gets the correct message bit-pair and this case arises with probability $1/4$. In that case, if the joint measurement result is $\ket{\Phi^+}$, then $\mathcal{A}$ can not be detected, but if the joint measurement result is $\ket{\Psi^+}$, then they can detect $\mathcal{A}$. The details are given in Table~\ref{table_attack}.
	
	\begin{table}[!htbp]
		\centering
		\caption{Different cases of intercept-and-resend attack on MDI-QD.}
		\resizebox{\textwidth}{!}{
			\begin{tabular}{|c|c|c|c|c|c|c|c|c|c|c|c|c|}
				\hline
				$k_i$ & ${k'}_i$ & $a_i$ & $b_i$ & $Q_{A_i}$ & $Q_{B_i}$ & ${Q'}_{A_i}$ & ${Q'}_{B_i}$ & \multicolumn{4}{|c|}{Prob. of joint measurement result} & Remark\\
				
				&     &  &   & &     &  &   & $\ket{\Phi^+}$ & $\ket{\Phi^-}$ & $\ket{\Psi^+}$ & $\ket{\Psi^-}$ &\\
				\hline
				$0$ & $1$ & $0$ & $0$  & $\ket{0}$  &  $\ket{0}$ & $\ket{+}$  &  $\ket{+}$ & $1/2$  & $0$ & $\mathbf{1/2}$  & $0$  & with probability\\
				&  & &&&& $\ket{+}$  &  $\ket{-}$ &  $0$ & $1/2$ & $0$  & $\mathbf{1/2}$ &$1/2$ cheating  \\
				&  & &&&& $\ket{-}$  &  $\ket{+}$ &  $0$& $1/2$ & $0$  & $\mathbf{1/2}$ &can be \\
				&  & &&&& $\ket{-}$  &  $\ket{-}$ & $1/2$ & $0$ & $\mathbf{1/2}$ & $0$ & detected  \\
				\hline
			\end{tabular}
		}
		\\
		\begin{flushleft}
			\textit{\begin{tiny}
					*Bold numbers denote the probabilities that errors have occurred.
			\end{tiny}}
		\end{flushleft}
		
		\label{table_attack}
	\end{table}
	Thus, in the case of the intercept-and -resend attack, \\
	$\Pr($cheating detected in $i$-th bit $ )$ = $\Pr($cheating detected in $i$-th bit $|k_i={k'}_i ) \Pr(k_i={k'}_i)+ \Pr($cheating detected in $i$-th bit $|k_i\neq {k'}_i ) \Pr(k_i \neq {k'}_i)= 0+ 1/2\times 1/2 =1/4$. Therefore, with probability $3/4$, $\mathcal{A}$ can do the attack without being detected.
	
	$\Pr(\mathcal{A}$ gets the exact $i$-th bit message pair $)= 1/2+1/2\times1/4= 5/8$, whereas $\Pr(\mathcal{A}$ guesses the exact $i$-th bit message pair randomly$)=1/4$. 
	
	To avoid this attack, we have modified the previous MDI-QD protocol by introducing an extra error estimation phase before the UTP jointly measures the qubits.
	
	\subsection{Our Proposed Modification}
	\label{mod qd}
	Steps 1, 2, 3 are the same as before in the MDI-QD protocol of Figure~\ref{algo:1qd}.
	\begin{enumerate}
		\setcounter{enumi}{3}
		\item Alice and Bob choose some random permutation and apply those on their respective sequences of qubits $Q_A$ and $Q_B$ and get new sequences of qubits $q_A$ and $q_B$.
		\item They send the prepared qubits $q_A $ and $q_B$ to a UTP.
		\item Alice and Bob randomly choose $\delta n$ number of common positions on sequences $Q_A$ and $Q_B$ to estimate the error in the channel, where $\delta \ll 1$ is a small fraction. Corresponding to these rounds, they do the followings: \label{2party_error1}
		\begin{enumerate}
			\item Each participant tells the positions and preparation bases of those qubits for those rounds to the UTP. 
			\item The UTP measures each single-qubit state in proper basis and announces the results.
			\item They reveal their respective qubits for these rounds and compare them with the results announced by the UTP.
			\item If the estimated error is greater than some predefined threshold value, then they abort. Else they continue and go to the next step.
		\end{enumerate}
		\item The UTP asks Alice and Bob the permutations which they have applied to their sequences.
		\item The UTP applies the inverse permutations, corresponding to the permutations chosen by Alice and Bob, on $q_A$ and $q_B$ to get $Q_A$ and $Q_B$ respectively.
		
		\item They discard the qubits corresponding to the above $\delta n$ positions. Their remaining sequence of prepared qubits are relabeled as $Q_A=\{Q_A[i]\}_{i=1}^{m}$ and $Q_B=\{Q_B[i]\}_{i=1}^{m}$, where $m=(1-\delta) n$. 
		\item They update their $n$-bit key to an $m$-bit key by discarding $\delta n$ number of key bits corresponding to the above $\delta n$ rounds. The updated key is relabeled as $k=k_1k_2\ldots k_{m}$.
	\end{enumerate}
	Then they follow Step 5 to Step 11 of the MDI-QD protocol in Figure~\ref{algo:1qd}.
	
	In this modified protocol, since Alice and Bob apply random permutations on their respective sequences of qubits before sending those qubits to the UTP and since those permutations are announced only after the error estimation phase is passed, at the time of sending those sequences $\mathcal{A}$ can not just guess a key bit and measure the qubits. Even if she gets some of the key bits, she can not guess the corresponding bases for sequences of qubits $q_A,~q_B$. Alice and Bob randomly choose $\delta n$ number of rounds to estimate the error in the channel (Step~\ref{2party_error1} of the modified protocol), where $\delta \ll 1$ is a small fraction. Corresponding to these rounds, they tell the key bits to the UTP and he measures each single-qubit state in proper basis and announces the results. Alice and Bob reveal their respective qubits for these rounds and compare them with the results announced by the UTP. 
	
	Let $\mathcal{A}$ intercept the sequences $q_A,~q_B$, measure those qubits and resend the sequences $q_A',~q_B'$. Let the $i$-th qubit pair be  $(q_{A_i},q_{B_i})$, which is prepared in the basis $(\mathcal{B}_{A_i},\mathcal{B}_{B_i})$, and suppose $\mathcal{A}$ independently chooses two bases $\mathcal{B}_{A_i}'$ and $\mathcal{B}_{B_i}'$ to measure $q_{A_i}$ and $q_{B_i}$, since they are not dependent on the $i$-th key bit. After measurement, let the state of the qubit pair be $(q_{A_i}',q_{B_i}')$. At the time of security checking, UTP measures $(q_{A_i}',q_{B_i}')$ in $(\mathcal{B}_{A_i},\mathcal{B}_{B_i})$ and gets the result $(q_{A_i}'',q_{B_i}'')$. 
		Thus the winning probability of $\mathcal{A}$ is 
		\begin{equation*} \label{eq-pr1}
		\begin{split}
		 &\Pr(q_{A_i}''=q_{A_i},~q_{B_i}''=q_{B_i}) \\
		&=\Pr(q_{A_i}''=q_{A_i})\Pr(q_{B_i}''=q_{B_i}) \\
		& = \{ \Pr(q_{A_i}''=q_{A_i}|~\mathcal{B}_{A_i} = \mathcal{B}_{A_i}')\Pr(\mathcal{B}_{A_i} = \mathcal{B}_{A_i}') + \Pr(q_{A_i}''=q_{A_i}|~\mathcal{B}_{A_i} \neq  \mathcal{B}_{A_i}') \Pr(\mathcal{B}_{A_i} \neq \mathcal{B}_{A_i}')\}\times \\
		& ~~~~ \{ \Pr(q_{B_i}''=q_{B_i}|~\mathcal{B}_{B_i} = \mathcal{B}_{B_i}')\Pr(\mathcal{B}_{B_i} = \mathcal{B}_{B_i}') + \Pr(q_{B_i}''=q_{B_i}|~\mathcal{B}_{B_i} \neq  \mathcal{B}_{B_i}') \Pr(\mathcal{B}_{B_i} \neq \mathcal{B}_{B_i}')\} \\
		&= \left[ \frac{1}{2}\{\Pr(q_{A_i}''=q_{A_i}|~\mathcal{B}_{A_i} = \mathcal{B}_{A_i}') + \Pr(q_{A_i}''=q_{A_i}|~\mathcal{B}_{A_i} \neq  \mathcal{B}_{A_i}')\}\right] \times \\
		& ~~~~~~~~~~ \left[ \frac{1}{2}\{ \Pr(q_{B_i}''=q_{B_i}|~\mathcal{B}_{B_i} = \mathcal{B}_{B_i}') + \Pr(q_{B_i}''=q_{B_i}|~\mathcal{B}_{B_i} \neq  \mathcal{B}_{B_i}')\}\right] \\
		&=  \frac{1}{4}\left(1+\frac{1}{2} \right)\left(1+\frac{1}{2} \right) =\frac{9}{16}.
		\end{split}
		\end{equation*}	
		Since Alice and Bob apply random permutations on their sequences $Q_A$ and $Q_B$, so from the measurement results, $\mathcal{A}$ can not get any information about the $i$-th bit pair of the secret message. The probability of getting the $i$-th bit pair is $1/4$ by randomly guessing the bits. However the detection probability of $\mathcal{A}$ is $1-\left( \frac{9}{16}\right)^{\delta n} $ and in this case Alice and Bob abort the protocol. 
	
	Table~\ref{table:comarison} compares the probabilities of relevant events between the MDI-QD~\cite{qip/Maitra17} and its modified version.
	
	\begin{table}[h]
		\caption{Comparison between the MDI-QD~\cite{qip/Maitra17} and its modified version.}
		\label{table:comarison}
		\renewcommand*{\arraystretch}{1.8}
		\resizebox{1.0\textwidth}{!}{
			\begin{tabular}{|c|c|c|}
				\hline
				\textbf{Probability of the event}                                                        & \textbf{MDI-QD~\cite{qip/Maitra17}} & \textbf{Our modified MDI-QD} \\
				\hline
				$\mathcal{A}$ gets the $i$-th bit pair                            & $5/8$  & $1/4$             \\
				\hline
				Alice, Bob can not detect $\mathcal{A}$ for the $i$-th measurement & {$3/4$}  & $9/16$              \\
				\hline
				Alice, Bob detect $\mathcal{A}$ &   $1-(3/4)^{\delta n}$      &    $1-(9/16)^{\delta n}$  \\
				\hline           
		\end{tabular}}
	\end{table}
	
	\section{Three Party Quantum Conference } \label{sec3}
	We extend the above QD protocol from two to three parties, thus leading to a protocol of quantum conference. Our proposed conference protocol is divided into two parts. Let Alice, Bob and Charlie be three participants of the conference. Also let Alice's, Bob's and Charlie's $m$ bit messages be $a$, $b$ and $c$ respectively, where $a=a_1a_2\ldots a_m$, $b=b_1b_2\ldots b_m$ and $c=c_1c_2\ldots c_m$.
	
	In the first part, Alice, Bob, and Charlie perform a Multi-party QKD protocol~\cite{matsumoto2007multiparty} to establish a secret key $k=k_1k_2\ldots k_m$ of $m$ bits between themselves. Then each of them uses the key to encode one's own message $M$ into the corresponding state $Q$, according to Subroutine~1. The details of the three party quantum conference protocol are given in Protocol~1.
	\sbline
	\begin{subroutine}{Message Encoding Strategy for Three Party Quantum Conference }
		\textbf{Inputs:} Own message ${M}=M_1M_2\ldots M_m$; key $k=k_1k_2\ldots k_m$.
		\sbline
		\textbf{Output:} Sequence of qubits $Q=Q_1Q_2\ldots Q_m$.
		\sbline
		\textit{The subroutine:}\\
		For $1 \leqslant i \leqslant m,$
		\begin{enumerate}
			\item if $M_i= 0$ and $k_i = 0$, prepares $Q_i=\ket{0}$.
			\item if $M_i= 1$ and $k_i = 0$, prepares $Q_i=\ket{1}$.
			\item if $M_i= 0$ and $k_i = 1$, prepares $Q_i=\ket{+}$.
			\item if $M_i= 1$ and $k_i = 1$, prepares $Q_i=\ket{-}$.
		\end{enumerate}	
		\label{algo:enc_conf}			
	\end{subroutine}
	
	\subsection{Protocol 1: Three Party Quantum Conference}
	\label{conf}
	The steps of the protocol is as follows:
	\begin{enumerate}
		\item 
		Alice, Bob and Charlie perform any multi-party QKD protocol (e.g.,~\cite{matsumoto2007multiparty}) to establish an $m$-bit secret key $k=k_1k_2\ldots k_m$ between themselves.
		
		\item Let the $m$-bit messages of Alice, Bob and Charlie be $a$, $b$ and $c$ respectively, where $a=a_1a_2\ldots a_m$, $b=b_1b_2\ldots b_m$ and $c=c_1c_2\ldots c_m$.
		
		\item For $1\leqslant i \leqslant m$, Alice, Bob and Charlie prepare the sequences of qubits $Q_A=\{Q_A[i]\}_{i=1}^{m}=({Q_A}_1,{Q_A}_2,\ldots,{Q_A}_m), Q_B=\{Q_B[i]\}_{i=1}^{m}=({Q_B}_1,{Q_B}_2,\ldots,{Q_B}_m)$ and $Q_C=\{Q_C[i]\}_{i=1}^{m}=({Q_C}_1,{Q_C}_2,\ldots,{Q_C}_m)$ respectively at their end by using Subroutine 1. 
		
		\item Alice, Bob, and Charlie choose some random permutation and apply those on their respective sequences of qubits $Q_A,Q_B$, and $Q_C$ and get new sequences of qubits $q_A,q_B$ and $q_C$.
		
		\item They send the prepared sequences of qubits $q_A ,q_B$, and $q_C$ to an untrusted fourth party (UFP). 
		
		\item Alice, Bob, and Charlie randomly choose $\delta m$ number of common positions on sequences $Q_A ,Q_B$ and $Q_C$ to estimate the error in the channel, where $\delta \ll 1$ is a small fraction. Corresponding to these $\delta m$ rounds, they do the following: \label{3party_error1}
		\begin{enumerate}
			\item Each participant tells the positions and preparation bases of those qubits for those rounds to the UFP. 
			\item The UFP  measures each single-qubit state in proper basis and announces the results.
			\item They reveal their respective qubits for these rounds and compare them with the results announced by the UFP.
			\item If the estimated error is greater than some predefined threshold value, then they abort. Else they continue and go to the next step.
		\end{enumerate}
		\item The UFP  asks Alice, Bob, and Charlie to tell the permutations which they have applied to their sequences.
		\item The UFP  applies the inverse permutations, corresponding to the permutations chosen by Alice, Bob, and Charlie, on $q_A,q_B$, and $q_C$ to get $Q_A ,Q_B$ and $Q_C$ respectively.
		
		\item They discard the qubits corresponding to the above $\delta m$ positions. Their remaining sequence of prepared qubits are relabeled as $Q_A=\{Q_A[i]\}_{i=1}^{m'}$, $Q_B=\{Q_B[i]\}_{i=1}^{m'}$ and $Q_C=\{Q_C[i]\}_{i=1}^{m'}$, where $m'=(1-\delta) m$.
		
		\item They update their $m$-bit key to an $m'$-bit key by discarding $\delta m$ number of key bits corresponding to the above $\delta m$ rounds. The updated key is relabeled as $k=k_1k_2\ldots k_{m'}$.
		\item  For $1\leqslant i \leqslant m'$, the UFP  measures the each three qubits state $(Q_{A_i},Q_{B_i},Q_{C_i})$ in basis $\mathcal{B}_3$ and announces the result.
		\item Alice, Bob and Charlie make a finite sequence $\{\mathcal{M}[i]\}_{i=1}^{m'}$ containing the measurement results, i.e., for $1\leqslant i \leqslant m'$, $\mathcal{M}[i]\in \{\ket{\Phi_{0}^{+}},\ket{\Phi_{0}^{-}},\ket{\Phi_{1}^{+}},\ket{\Phi_{1}^{-}},\ket{\Phi_{2}^{+}}, \ket{\Phi_{2}^{-}}, \ket{\Phi_{3}^{+}}, \ket{\Phi_{3}^{-}}\}$ is the $i$-th measurement result announced by the UFP .
		
		\item They randomly choose $\gamma m'$ number of measurement results $\mathcal{M}[i]$ from the sequence $\{\mathcal{M}[i]\}_{i=1}^{m'}$ to estimate the error (may be introduced by the UFP ), where $\gamma \ll 1$ is a small fraction. \label{3party_error2}
		\begin{enumerate}
			\item They reveal their respective message bits for these rounds.
			
			\item If the estimated error is greater than some predefined threshold value, then they abort. Else they continue and go to the next step.\label{3prty_error_end}
			
		\end{enumerate}
		
		\item Their remaining sequence of measurement results is relabeled as $\{\mathcal{M}[i]\}_{i=1}^{n}$, where $n=(1-\gamma) m'$.
		
		\item They update their $m'$-bit key to an $n$-bit key by discarding $\gamma m'$ number of key bits corresponding to the above $\gamma m'$ rounds. The updated key is relabeled as $k=k_1k_2\ldots k_{n}$.
		
		\item Each of Alice, Bob, and Charlie applies Algorithm~\ref{3_Party_msg_recons} to get others' messages.
		
	\end{enumerate}
	Note that in this protocol, there are two error estimation phases. The first one checks if there is any adversary (other than the UFP ) in the channel who tries to get some information about the messages or change the messages. In this case, if the 1st error estimation phase does not pass, then Alice, Bob, and Charlie abort the protocol. Thus, in this step, the motivation of the UFP  being correct is that there is no information gain for him/her if the parties abort the protocol. The next error estimation phase is to check if there is any error introduced by the UFP . 
	
	\begin{algorithm}
		\setlength{\textfloatsep}{0.05cm}
		\setlength{\floatsep}{0.05cm}
		\KwIn{Own message , measurement results $\{\mathcal{M}[i]\}_{i=1}^{n}$, key $k$. }
		\KwOut{Others' messages.}
		
		\begin{enumerate}
			\item For $1\leqslant i \leqslant n$, if $k_i = 0$, then each participant can learn the $i$-th bit of others' messages from the measurement result $\mathcal{M}[i]$ and their own message (see Table-\ref{conf_table}).
			\item For $1\leqslant i \leqslant n$, if  $k_i = 1$, then from the measurement result $\mathcal{M}[i]$ and their own message each participant can learn the $i$-th bit of others messages are same or different (see Table-\ref{conf_table}). Let $c=wt(k)$. \label{msg_info}
			\begin{enumerate}
				\item Alice, Bob and Charlie prepare ordered sets of qubits $S_A$, $S_B$ and $S_C$ respectively, corresponding to their message bit where the key bit is $1$. They prepare the qubits at their end according to the following strategy. Each of $S_A$, $S_B$ and $S_C$ contain $c$ number of qubits. For $1\leqslant j \leqslant c$ and if $k_i=1$ is the $j$-th $1$ in $k$, then
				\begin{itemize} \label{2nd encode}
					\item if $a_i$ ($b_i,c_i$)$ = 0$ and $i$ is even, prepares $S_A[j] ~(S_B[j],~S_C[j])~=\ket{0}$.
					\item  if $a_i$ ($b_i,c_i$)$ = 1$ and $i$ is even, prepares $S_A[j] ~(S_B[j],~S_C[j])~=\ket{1}$.
					\item if $a_i$ ($b_i,c_i$)$ = 0$ and $i$ is odd, prepares $S_A[j] ~(S_B[j],~S_C[j])~=\ket{+}$.
					\item  if $a_i$ ($b_i,c_i$)$ = 1$ and $i$ is odd, prepares $S_A[j] ~(S_B[j],~S_C[j])~=\ket{-}$.
				\end{itemize}
				\item Alice, Bob and Charlie prepare sets of $d$ decoy photons $D_A$, $D_B$ and $D_C$ respectively, where the decoy photons are randomly chosen from $\{\ket{0},\ket{1},\ket{+},\ket{-}\}$. They randomly insert their decoy photons into their prepared qubits sets and make new ordered sets $S_A'$, $S_B'$ and $S_C'$. They also choose random permutations $R_A$, $R_B$, $R_C$ and apply those on their respective sets $S_A'$, $S_B'$, $S_C'$ to get the sets $S_A''$, $S_B''$, $S_C''$ respectively.
				\item  Each of them sends its set to the next participant in a circular way. That is, Alice sends $S_A''$ to Bob, who sends $S_B''$ to Charlie, who in turn sends $S_C''$ to Alice.
				\item After receiving the qubits from the previous participant, each of them announces the random permutations and the positions, states of their decoy photons.
				\item They apply the inverse permutations and verify the decoy photons to check eavesdropping. If there exists any eavesdropper in the quantum channel, they abort the protocol, else they go to the next step.\label{3party_error3}
				\item Now everyone knows the basis of the qubits of $S_A$, $S_B$ and $S_C$. So they can measure those qubits to get the exact message bits of the previous participant from whom they got those qubits. \label{rcv_qubit}
				
			\end{enumerate} 
			
		\end{enumerate}
		
		\caption{Three Party Message Reconstruction Algorithm.}
		\label{3_Party_msg_recons}
		\setlength{\textfloatsep}{0.05cm}
		\setlength{\floatsep}{0.05cm}
	\end{algorithm}

	\begin{sidewaystable}[]
		\centering
		\renewcommand*{\arraystretch}{1.8}
		\caption{Different cases in the three party quantum conference.}
		\setlength{\tabcolsep}{8pt}
		\resizebox{1.0\textwidth}{!}{
			\begin{tabular}{|c|c|c|c|c|c|c|c|c|c|c|c|c|c|}
				\hline
				\multicolumn{3}{|c|}{Bits to Communicate} &  \multicolumn{3}{|c|}{ Qubits prepared by } & \multicolumn{8}{|c|}{Probabilities of measurement results $\mathcal{M}[i]$ at UFP's end}\\
				\hline
				{} Alice  & Bob & Charlie  &  Alice (${Q_A}_i$) &  Bob (${Q_B}_i$) & Charlie (${Q_C}_i$)  & $\ket{\Phi_0^+}$ & $\ket{\Phi_0^-}$ & $\ket{\Phi_1^+}$ & $\ket{\Phi_1^-}$ & $\ket{\Phi_2^+}$ & $\ket{\Phi_2^-}$ & $\ket{\Phi_3^+}$& $\ket{\Phi_3^-}$ \\
				\hline
				$0$ & $0$ & $0$ & $\ket{0}$  &  $\ket{0}$ &  $\ket{0}$ & $1/2$ & $1/2$  & $0$ & $0$ &$0$ & $0$ & $0$ & $0$ \\
				$0$ & $0$ & $1$ & $\ket{0}$  &  $\ket{0}$ &  $\ket{1}$ &  $0$ & $0$ & $1/2$ & $1/2$ & $0$ & $0$ & $0$  & $0$ \\
				$0$  & $1$ & $0$ & $\ket{0}$  &  $\ket{1}$ &  $\ket{0}$ &  $0$ & $0$ & $0$ & $0$ & $1/2$ & $1/2$ & $0$ & $0$ \\
				$0$ & $1$ & $1$ & $\ket{0}$  &  $\ket{1}$ &  $\ket{1}$  & $0$ & $0$ & $0$ & $0$ & $0$ & $0$ & $1/2$ & $1/2$ \\
				$1$ & $0$ & $0$ & $\ket{1}$  &  $\ket{0}$ &  $\ket{0}$ & $0$ & $0$ & $0$ & $0$ & $0$ & $0$ & $1/2$ & $1/2$ \\
				$1$  & $0$ & $1$ & $\ket{1}$  &  $\ket{0}$ &  $\ket{1}$ & $0$ & $0$ & $0$ & $0$ & $1/2$ & $1/2$ & $0$ & $0$ \\
				$1$ & $1$ & $0$ & $\ket{1}$  &  $\ket{1}$ &  $\ket{0}$ &  $0$ & $0$ & $1/2$ & $1/2$ & $0$ & $0$ & $0$  & $0$ \\
				$1$ & $1$ & $1$ & $\ket{1}$  &  $\ket{1}$ &  $\ket{1}$ & $1/2$ & $1/2$  & $0$ & $0$ & $0$ & $0$ & $0$ & $0$ \\
				\hline
				$0$ & $0$ & $0$ & $\ket{+}$  &  $\ket{+}$  &  $\ket{+}$ & $1/4$ & $0$  & $1/4$ & $0$ & $1/4$ & $0$ & $1/4$ & $0$ \\
				$0$ & $0$ & $1$ & $\ket{+}$  &  $\ket{+}$  &  $\ket{-}$ & $0$ & $1/4$  &  $0$ & $1/4$ & $0$  & $1/4$ &  $0$  & $1/4$ \\
				$0$  & $1$ & $0$ & $\ket{+}$  &  $\ket{-}$  &  $\ket{+}$ & $0$ & $1/4$  &  $0$ & $1/4$ & $0$  & $1/4$ &  $0$  & $1/4$ \\
				$0$ & $1$ & $1$ & $\ket{+}$  &  $\ket{-}$  &  $\ket{-}$ & $1/4$ & $0$  & $1/4$ & $0$ & $1/4$ & $0$ & $1/4$ & $0$ \\
				$1$ & $0$ & $0$ & $\ket{-}$  &  $\ket{+}$  &  $\ket{+}$ & $0$ & $1/4$  &  $0$ & $1/4$ & $0$  & $1/4$ &  $0$  & $1/4$ \\
				$1$  & $0$ & $1$ &  $\ket{-}$  &  $\ket{+}$  &  $\ket{-}$ & $1/4$ & $0$  & $1/4$ & $0$ & $1/4$ & $0$ & $1/4$ & $0$  \\
				$1$ & $1$ & $0$ & $\ket{-}$  &  $\ket{-}$  &  $\ket{+}$ & $1/4$ & $0$  & $1/4$ & $0$ & $1/4$ & $0$ & $1/4$ & $0$  \\
				$1$ & $1$ & $1$ & $\ket{-}$  &  $\ket{-}$  &  $\ket{-}$  & $0$ & $1/4$  &  $0$ & $1/4$ & $0$  & $1/4$ &  $0$  & $1/4$\\
				\hline
		\end{tabular}}
		\label{conf_table}
		`  \end{sidewaystable}
	
	\subsection{Correctness of Three Party Quantum Conference Protocol}
	In our proposed protocol, Alice, Bob and Charlie first prepare qubits corresponding to their messages and shared key and then send those qubits to the fourth party (UFP). After that, UFP  measures each of the three qubits state (one from Alice, one from Bob and one from Charlie) in basis $\mathcal{B}_3=\{\ket{\Phi_{0}^{+}},\ket{\Phi_{0}^{-}},\ket{\Phi_{1}^{+}},\ket{\Phi_{1}^{-}},\ket{\Phi_{2}^{+}}, \ket{\Phi_{2}^{-}}, \ket{\Phi_{3}^{+}}, \ket{\Phi_{3}^{-}}\}$ and announces the result. Now, we can say the following from Table~\ref{conf_table}:
	\begin{itemize}
		\item If the prepared qubit of Alice is $\ket{0}$($\ket{1})$, then Alice guesses message bit of Bob and Charlie ($b_i$ and $c_i$) with probability $1$ as follows:
		\begin{equation*}
		\text{Measurement result} =
		\begin{cases}
		\ket{\Phi_{0}^{+}}$ or $\ket{\Phi_{0}^{-}} \Rightarrow & b_i=0(1) \text{ and }c_i=0(1);\\
		\ket{\Phi_{1}^{+}}$ or $\ket{\Phi_{1}^{-}} \Rightarrow & b_i=0(1) \text{ and }c_i=1(0);\\
		\ket{\Phi_{2}^{+}}$ or $\ket{\Phi_{2}^{-}} \Rightarrow & b_i=1(0) \text{ and }c_i=0(1);\\
		\ket{\Phi_{3}^{+}}$ or $\ket{\Phi_{3}^{-}} \Rightarrow & b_i=1(0) \text{ and }c_i=1(0).
		\end{cases}
		\end{equation*}
		
		\item If the prepared qubit of Alice is $\ket{+}$($\ket{-})$, then Alice guesses the XOR function of message bits of Bob and Charlie with probability $1$ as follows:
		\begin{equation*}
		\text{Measurement result} =
		\begin{cases}
		\ket{\Phi_{0}^{+}}$ or $\ket{\Phi_{1}^{+}}$ or $\ket{\Phi_{2}^{+}}$ or $\ket{\Phi_{3}^{+}}  \Rightarrow & b_i \oplus c_i =0(1);\\
		\ket{\Phi_{0}^{-}}$ or $\ket{\Phi_{1}^{-}}$ or $\ket{\Phi_{2}^{-}}$ or $\ket{\Phi_{3}^{-}}  \Rightarrow & b_i \oplus c_i =1(0).
		\end{cases}
		\end{equation*}
		In this case, Charlie sends her encoded qubit to Alice (the encoding process is given in Step~\ref{2nd encode} of Algorithm~\ref{3_Party_msg_recons}). Since Alice knows the basis of the received qubit from Charlie, by measuring the qubit in the proper basis, Alice can know the message bit $c_i$ of Charlie. Then from $b_i \oplus c_i$, she can get $b_i$ also.
	\end{itemize}
	
	A similar thing happens for Bob and Charlie too.     
	From the above discussion, we see that for all the cases Alice, Bob, and Charlie can conclude the communicated bit of the other parties with probability $1$. Hence our protocol is giving the correct result.
	
	\subsection{Security Analysis of the Three Party Quantum Conference Protocol}
	In this section, we discuss the security of our proposed three-party quantum conference protocol against the common known attacks which $\mathcal{A}$ can adopt. If there exists some adversary in the channel and the legitimate parties can detect her with a non-negligible probability, then we call our protocol as secure.
	
	We first show that if the UFP does some cheating, it can be detected by the players at the error estimation phase of the protocol (Step~\ref{3party_error2} of Protocol 1).    
	\begin{sidewaystable}[]
		\centering
		\renewcommand*{\arraystretch}{1.8}
		\caption{Different cases when UFP is dishonest in the three party quantum conference.}
		\setlength{\tabcolsep}{8pt}
		\resizebox{1.0\textwidth}{!}{
			\begin{tabular}{|c|c|c|c|c|c|c|c|c|c|c|c|}
				\hline
				{UFP choses} &  \multicolumn{3}{|c|}{UFP's measurement results} & \multicolumn{8}{|c|}{Probability that UFP guesses $\mathcal{M}'[i]$}\\
				\hline
				{} measurement basis  &  Alice (${Q_A}'_i$) &  Bob (${Q_B}'_i$) & Charlie (${Q_C}'_i$)  & $\ket{\Phi_0^+}$ & $\ket{\Phi_0^-}$ & $\ket{\Phi_1^+}$ & $\ket{\Phi_1^-}$ & $\ket{\Phi_2^+}$ & $\ket{\Phi_2^-}$ & $\ket{\Phi_3^+}$& $\ket{\Phi_3^-}$ \\
				\hline
				\multirow{8}{*}{$Z$} & $\ket{0}$  &  $\ket{0}$ &  $\ket{0}$ & $1/2$ & $1/2$  & $0$ & $0$ &$0$ & $0$ & $0$ & $0$ \\
				& $\ket{0}$  &  $\ket{0}$ &  $\ket{1}$ &  $0$ & $0$ & $1/2$ & $1/2$ & $0$ & $0$ & $0$  & $0$ \\
				& $\ket{0}$  &  $\ket{1}$ &  $\ket{0}$ &  $0$ & $0$ & $0$ & $0$ & $1/2$ & $1/2$ & $0$ & $0$ \\
				& $\ket{0}$  &  $\ket{1}$ &  $\ket{1}$  & $0$ & $0$ & $0$ & $0$ & $0$ & $0$ & $1/2$ & $1/2$ \\
				& $\ket{1}$  &  $\ket{0}$ &  $\ket{0}$ & $0$ & $0$ & $0$ & $0$ & $0$ & $0$ & $1/2$ & $1/2$ \\
				& $\ket{1}$  &  $\ket{0}$ &  $\ket{1}$ & $0$ & $0$ & $0$ & $0$ & $1/2$ & $1/2$ & $0$ & $0$ \\
				& $\ket{1}$  &  $\ket{1}$ &  $\ket{0}$ &  $0$ & $0$ & $1/2$ & $1/2$ & $0$ & $0$ & $0$  & $0$ \\
				& $\ket{1}$  &  $\ket{1}$ &  $\ket{1}$ & $1/2$ & $1/2$  & $0$ & $0$ & $0$ & $0$ & $0$ & $0$ \\
				\hline
				\multirow{8}{*}{$X$} & $\ket{+}$  &  $\ket{+}$  &  $\ket{+}$ & $1/4$ & $0$  & $1/4$ & $0$ & $1/4$ & $0$ & $1/4$ & $0$ \\
				& $\ket{+}$  &  $\ket{+}$  &  $\ket{-}$ & $0$ & $1/4$  &  $0$ & $1/4$ & $0$  & $1/4$ &  $0$  & $1/4$ \\
				& $\ket{+}$  &  $\ket{-}$  &  $\ket{+}$ & $0$ & $1/4$  &  $0$ & $1/4$ & $0$  & $1/4$ &  $0$  & $1/4$ \\
				& $\ket{+}$  &  $\ket{-}$  &  $\ket{-}$ & $1/4$ & $0$  & $1/4$ & $0$ & $1/4$ & $0$ & $1/4$ & $0$ \\
				& $\ket{-}$  &  $\ket{+}$  &  $\ket{+}$ & $0$ & $1/4$  &  $0$ & $1/4$ & $0$  & $1/4$ &  $0$  & $1/4$ \\
				&  $\ket{-}$  &  $\ket{+}$  &  $\ket{-}$ & $1/4$ & $0$  & $1/4$ & $0$ & $1/4$ & $0$ & $1/4$ & $0$  \\
				& $\ket{-}$  &  $\ket{-}$  &  $\ket{+}$ & $1/4$ & $0$  & $1/4$ & $0$ & $1/4$ & $0$ & $1/4$ & $0$  \\
				& $\ket{-}$  &  $\ket{-}$  &  $\ket{-}$  & $0$ & $1/4$  &  $0$ & $1/4$ & $0$  & $1/4$ &  $0$  & $1/4$\\
				\hline
		\end{tabular}}
		\label{ufp_conf_table}
	\end{sidewaystable} 
	Let UFP measure each of the three qubits $Q_{A_i},Q_{B_i},Q_{C_i}$ in a randomly chosen basis ($Z$ or $X$) instead of measuring $(Q_{A_i},Q_{B_i},Q_{C_i})$ in $\mathcal{B}_3$ basis. Now UFP checks the individual measurement results and decides to announce an $\mathcal{M}'[i]\in \{\ket{\Phi_{0}^{+}},\ket{\Phi_{0}^{-}},\ket{\Phi_{1}^{+}},\ket{\Phi_{1}^{-}},\ket{\Phi_{2}^{+}}, \ket{\Phi_{2}^{-}}, \ket{\Phi_{3}^{+}}, \ket{\Phi_{3}^{-}}\}$ corresponding to the states which can arrive if he measures in the correct basis (see Table~\ref{ufp_conf_table}). For example, if UFP measures in $Z$-basis and gets the result $\ket{0}\ket{0}\ket{1}$ then he announces $\mathcal{M}'[i]$ from the set $\{\ket{\Phi_{1}^{+}},\ket{\Phi_{1}^{-}}\}$. Again if he measures in $X$-basis and gets the result $\ket{-}\ket{+}\ket{+}$ then he announces $\mathcal{M}'[i]$ from the set $\{\ket{\Phi_{0}^{-}},\ket{\Phi_{1}^{-}},\ket{\Phi_{2}^{-}},\ket{\Phi_{3}^{-}}\}$.
	\\We now calculate the winning probability $p$ of UFP for correctly guessing the $i$-th measurement result $\mathcal{M}[i]$. Let the preparation basis for the initial qubits $Q_{A_i},Q_{B_i},Q_{C_i}$ be $\mathcal{B}$ and UFP chooses the basis $\mathcal{B}'$.  Then we have,
	\begin{equation*} \label{eq-UFP-3party}
		\begin{split}
		p &=\Pr(\mathcal{M}'[i]=\mathcal{M}[i]) \\
		& = \Pr(\mathcal{M}'[i]=\mathcal{M}[i] |~\mathcal{B} = \mathcal{B}')\Pr(\mathcal{B} = \mathcal{B}') + \Pr(\mathcal{M}'[i]=\mathcal{M}[i] |~\mathcal{B} \neq \mathcal{B}')\Pr(\mathcal{B} \neq \mathcal{B}')\\
		&= \frac{1}{2}\{\Pr(\mathcal{M}'[i]=\mathcal{M}[i] |~\mathcal{B} = \mathcal{B}') + \Pr(\mathcal{M}'[i]=\mathcal{M}[i] |~\mathcal{B} \neq \mathcal{B}')\}\\
		&= \frac{1}{2}\{\Pr(\mathcal{M}'[i]=\mathcal{M}[i] |~\mathcal{B} = \mathcal{B}') + \Pr(\mathcal{M}'[i]=\mathcal{M}[i] |~\mathcal{B}=X, \mathcal{B}'=Z)+ \\ &~~~~~~~~~~~~~~~~~~~~~~~~~~~\Pr(\mathcal{M}'[i]=\mathcal{M}[i] |~\mathcal{B}=Z, \mathcal{B}'=X)\}\\
		&=  \frac{1}{2}\left(1+\frac{1}{2}+\frac{1}{4} \right) =\frac{7}{8}.
		\end{split}
	\end{equation*}
	Therefore the legitimate parties can detect this eavesdropping with probability $1-p^{\gamma m'}$, which is a non-negligible probability for large $\gamma m'$. 	
	
	Next, we consider four types of attacks (intercept-and-resend attack, entangle-and-measure attack, Denial-of-Service (DoS) attack, man-in-the-middle attack) and show that our protocol is secure against these attacks. 
		\begin{enumerate}
			\item \textbf{Intercept-and-resend attack}\\
			Here we consider the intercept-and-resend attack by an adversary $\mathcal{A}$ (other than the UFP). In this attack model, $\mathcal{A}$ intercepts the qubits from the quantum channel, then she measures those qubits and resends to the receiver. First let us assume that $\mathcal{A}$ intercepts $q_A$, measures the qubits in randomly chosen bases ($Z$ or $X$) and notes down the measurement results. Due to the measurements by $\mathcal{A}$, let the sequence $q_A$ changes to $q_A'$ and she resends $q_A'$ to UFP. After receiving the sequence $q_A'$, Alice tells UFP some random positions of the sent qubits and their preparation bases, then UFP measures those qubits and announces the results. Let the $i$-th qubit $q_{A_i}$ prepared in basis $\mathcal{B}_{A_i}$, and $\mathcal{A}$ chooses basis $\mathcal{B}_{A_i}'$ to measure $q_{A_i}$. At the time of security checking, UFP measures $q_{A_i}'$ in $\mathcal{B}_{A_i}$ and gets the result $q_{A_i}''$. 
			
			Thus the winning probability of $\mathcal{A}$ is 
			\begin{equation*} \label{eq-intercept-3party}
			\begin{split}
			p_1 &=\Pr(q_{A_i}''=q_{A_i}) \\
			& = \Pr(q_{A_i}''=q_{A_i}|~\mathcal{B}_{A_i} = \mathcal{B}_{A_i}')\Pr(\mathcal{B}_{A_i} = \mathcal{B}_{A_i}') + \Pr(q_{A_i}''=q_{A_i}|~\mathcal{B}_{A_i} \neq  \mathcal{B}_{A_i}') \Pr(\mathcal{B}_{A_i} \neq \mathcal{B}_{A_i}')\\
			&= \frac{1}{2}\{\Pr(q_{A_i}''=q_{A_i}|~\mathcal{B}_{A_i} = \mathcal{B}_{A_i}') + \Pr(q_{A_i}''=q_{A_i}|~\mathcal{B}_{A_i} \neq  \mathcal{B}_{A_i}')\}\\
			&=  \frac{1}{4}\left(1+\frac{1}{2} \right) =\frac{3}{4}.
			\end{split}
			\end{equation*}	
			Similarly, when $\mathcal{A}$ intercepts $q_B$ and $q_C$, then the winning probability of $\mathcal{A}$ is $p_2=\frac{3}{4}$ and $p_3=\frac{3}{4}$ respectively. Note that Alice, Bob, and Charlie apply random permutations on their respective sequences of qubits, and those permutations are announced only if the error estimation phase is passed after the qubits arrive at their destinations. So at the time of sending those sequences, $\mathcal{A}$ can not just guess a key bit and measure the qubits in the corresponding bases. Even if she gets some of the key bits, she can not guess the corresponding bases for sequences of qubits $q_A ,q_B$, $q_C$. Therefore measuring the qubits of $q_A ,q_B$, $q_C$ are independent events to $\mathcal{A}$ and thus the winning probability of $\mathcal{A}$ for this attack is $p_1p_2p_3=(\frac{3}{4})^3$. Alice, Bob, and Charlie randomly choose $\delta m$ number of rounds to estimate the error in the channel (Step~\ref{3party_error1} of Protocol 1), where $\delta \ll 1$ is a small fraction. Corresponding to these rounds, they tell the positions and preparation bases of the qubits to the UFP . Next, the UFP  measures each single qubit state in proper basis and announces the result. Alice, Bob, and Charlie reveal their respective qubits for these rounds and compare them with the results announced by UFP and calculate the error rate in the quantum channel. Thus the probability that they can detect the existence of $\mathcal{A}$ is $1-\left( \frac{3}{4}\right) ^{3\delta m}$, and in this case the legitimate parties terminate the protocol.\\
			Next we consider $\mathcal{A}$ tries to eavesdrop in the second phase of transmission of qubits (Step~\ref{msg_info} of  Algorithm~\ref{3_Party_msg_recons}). Suppose $\mathcal{A}$ intercepts the  sequences $S_A'',S_B'',S_C''$ from the quantum channel, measures them in $Z$ or $X$ basis and then resends those sequences to the receivers. Since each of $S_A'',S_B'',S_C''$ contains $d$ decoy photons, then these intermediate measurements change the states of those decoy photons. Let the $i$-th decoy photon of Alice be $D_{A_i}$ prepared in basis $\mathcal{B}$, where $\mathcal{B}=Z$ or $X$, and after $\mathcal{A}$ measures in $\mathcal{B}'$ basis the state becomes $D_{A_i}'$. When Alice announces the preparation basis of $D_{A_i}$, then Bob measures $D_{A_i}'$ in basis $\mathcal{B}$ and gets $D_{A_i}''$. We now calculate the probability that $D_{A_i}=D_{A_i}''$ as follows, 
			\begin{equation*} \label{eq-pr-intercept-2nd}
			\begin{split}
			&\Pr(D_{A_i}''=D_{A_i}) \\
			& =  \Pr(D_{A_i}''=D_{A_i}|~\mathcal{B} = \mathcal{B}')\Pr(\mathcal{B} = \mathcal{B}') + \Pr(D_{A_i}''=D_{A_i}|~\mathcal{B} \neq \mathcal{B}')\Pr(\mathcal{B} \neq \mathcal{B}') \\
			&= \frac{1}{2}[\Pr(D_{A_i}''=D_{A_i}|~\mathcal{B} = \mathcal{B}') + \Pr(D_{A_i}''=D_{A_i}|~\mathcal{B} \neq \mathcal{B}')] \\
			&=  \frac{1}{2}\left[1 + \frac{1}{2}\right]=\frac{3}{4}.
			\end{split}
			\end{equation*}
			Thus the probability that Alice and Bob can detect the existence of $\mathcal{A}$ is 
			$1-\left( \frac{3}{4}\right) ^d$, where $d$ is the number of decoy photon. Similarly for the other sequences of qubits.
			\item \textbf{Entangle-and-measure attack}\\
			Let us discuss another attack, called entangle-and-measure attack, by an adversary $\mathcal{A}$. For this attack, $\mathcal{A}$ does the following: when Alice sends her sequence of qubits $q_A$ to the UFP , then $\mathcal{A}$ takes each qubit $q_{A_i}$, $1 \leqslant i \leqslant m$, from the channel and takes an ancillary qubit $\ket{b}$, which is in state  $\ket{0}$, from her own. $\mathcal{A}$ applies a CNOT gate with control $q_{A_i}$ and target $\ket{b}$, and then she sends $q_{A_i}$ to the UFP . The joint state becomes $\ket{00}$, $\ket{11}$, $\ket{\Phi^+}$ and $\ket{\Phi^-}$, corresponding to the state of $q_{A_i}$, which are $\ket{0}$, $\ket{1}$, $\ket{+}$ and $\ket{-}$ respectively. Also $\mathcal{A}$ does the same thing with the qubits of Bob and Charlie. After the UFP  receives all the qubits, Alice, Bob and Charlie randomly choose $\delta m$ number of rounds to estimate the error in channel (Step~\ref{3party_error1} of Protocol 1), where $\delta \ll 1$ is a small fraction. Corresponding to these rounds, they tell the positions and preparation bases of the qubits to the UFP , who then measures each of the single qubit state in proper basis and announces the result. Alice, Bob and Charlie reveal their respective qubits for these rounds and compare with the results announced by the UFP. 
			
			Let UFP get the measurement result $q_{A_i}'$ by measuring the state $q_{A_i}$ prepared in basis $\mathcal{B}$. Now if the original state of $q_{A_i}$ is $\ket{0}$ or $\ket{1}$, then no error occurs. But if the original state of $q_{A_i}$ is $\ket{+}$ or $\ket{-}$, then an error will occur with probability $1/2$, as $\ket{\Phi^+}= \frac{1}{\sqrt{2}}(\ket{00}+ \ket{11})=\frac{1}{\sqrt{2}}(\ket{++}+ \ket{--})$ and $\ket{\Phi^{-}}=\frac{1}{\sqrt{2}}(\ket{00}- \ket{11})=\frac{1}{\sqrt{2}}(\ket{++}- \ket{--})$. Thus Alice, Bob and Charlie abort the protocol. Let us calculate the probability of the event $q_{A_i}'=q_{A_i}$.
			\begin{equation*} \label{eq-pr-entangled}
			\begin{split}
			p_1 &= \Pr(q_{A_i}'=q_{A_i}) \\
			& =  \Pr(q_{A_i}'=q_{A_i}|~\mathcal{B} = Z)\Pr(\mathcal{B} = Z) + \Pr(q_{A_i}'=q_{A_i}|~\mathcal{B} =X)\Pr(\mathcal{B} =X) \\
			&= \frac{1}{2}[q_{A_i}'=q_{A_i}|~\mathcal{B} = Z) + \Pr(q_{A_i}'=q_{A_i}|~\mathcal{B} =X)] \\
			&=  \frac{1}{2}\left[1 + \frac{1}{2}\right]=\frac{3}{4}.
			\end{split}
			\end{equation*}
			Similarly we can calculate $p'_2=\Pr(q_{B_i}'=q_{B_i})=\frac{3}{4}$, $p'_3=\Pr(q_{C_i}'=q_{C_i})=\frac{3}{4}$. Thus for $1 \leqslant i \leqslant m$, the winning probability of $\mathcal{A}$ is $p'_1p'_2p'_3=\left( \frac{3}{4}\right) ^3$ and the legitimate party can detect him at the time of security checking with probability $1-\left( \frac{3}{4}\right) ^{3\delta m}$. Similar argument follows for the second round of communication.
			\item \textbf{Denial-of-service (DoS) attack}\\
			In this attack model, $\mathcal{A}$ applies a random unitary operator $\mathcal{U} \neq I$ on the qubits to tamper the original message and introduce noise in the channel. 
			This attack can also be detected in the same way as discussed above. Let $\mathcal{U}=\sum_{j=1}^4 w_jP_j$, where $P_j$s are the Pauli matrices $I$, $\sigma_x$, $i\sigma_y$ and $\sigma_{z}$ for $1 \leq j \leq 4$ respectively~\cite{nielsen2002quantum}, and they form a basis for the space of all $2 \times 2$ Hermitian matrices. Since $\mathcal{U}$ is unitary, $\sum_{j=1}^4 w^2_j=1$. Now the winning probability of $\mathcal{A}$ is $p_4=\sum_{j=1}^4 h_jw^2_j$, where $h_j$s are the winning probabilities of $\mathcal{A}$ when she applies $P_j$s respectively. Thus $h_1=1$, $h_2=1/2$, $h_3=0$ and $h_4=1/2$ as $I$ does not change any state, $\sigma_x$ changes the states in $Z$-basis, $i\sigma_y$ changes the states in both $Z$-basis and $X$-basis, and $\sigma_z$ changes the states in $X$-basis. Hence in the security check process Alice, Bob and Charlie find this eavesdropping with probability $1-{p_4}^{3\delta m}>0$. Similarly for the second phase of communication, the legitimate parties can detect $\mathcal{A}$ with probability $1-{p_4}^{3d}>0$, where $d$ is the number of decoy states.
			\item \textbf{Man-in-the-middle attack}\\
			For this attack, $\mathcal{A}$ prepares three finite sequences of length $m$, of single qubit states $q_A',q_B'$ and $q_C'$, whose elements are randomly selected between $\ket{0}, \ket{1}, \ket{+}$ and $\ket{-}$. When Alice, Bob, and Charlie send their prepared sequences of qubits $q_A ,q_B$ and $q_C$ to the UFP , then $\mathcal{A}$ intercepts $q_A ,q_B$, $q_C$ and keeps those with her. Instead of $q_A ,q_B$ and $q_C$, she sends $q_A',q_B'$ and $q_C'$ to the UFP . Note that Alice, Bob, and Charlie apply random permutations on their respective sequences of qubits, and those permutations are announced only if the error estimation phase is passed after the qubits arrive at their destinations. So at the time of sending those sequences, $\mathcal{A}$ can not just guess a key bit and prepare her qubits. Even if she gets some of the key bits, she can not guess the corresponding bases for the sequences of qubits $q_A ,q_B$, $q_C$. Alice, Bob, and Charlie randomly choose $\delta m$ number of rounds to estimate the error in channel (Step~\ref{3party_error1} of Protocol 1), where $\delta \ll 1$ is a small fraction. Corresponding to these rounds, they tell the positions and preparation bases of the qubits to the UFP. Next, the UFP  measures each single qubit state in proper basis and announces the result. Alice, Bob, and Charlie reveal their respective qubits for these rounds and compare them with the results announced by UFP. Since the elements of $q_A',q_B'$, and $q_C'$ are randomly chosen by $\mathcal{A}$, thus they introduce error in the channel. Let us calculate the probability that Alice, Bob and Charlie can detect this eavesdropping and so they abort the protocol.\\
			For each $i$, let the $i$-th qubit of Alice be $q_{A_i}$ prepared in basis $\mathcal{B}_{A_i}$, and $\mathcal{A}$ prepare $q_{A_i}'$ in basis $\mathcal{B}_{A_i}'$. At the time of security checking, UFP measures $q_{A_i}'$ in $\mathcal{B}_{A_i}$ and gets the result $q_{A_i}''$. Now three cases may arise,
			\begin{itemize}
				\item If $\mathcal{B}_{A_i} = \mathcal{B}_{A_i}'$ and $q_{A_i}=q_{A_i}'$, then $q_{A_i}''=q_{A_i}$ with probability $1$.
				\item If $\mathcal{B}_{A_i} = \mathcal{B}_{A_i}'$ and $q_{A_i} \neq q_{A_i}'$, then $q_{A_i}''=q_{A_i}$ with probability $0$.
				\item If $\mathcal{B}_{A_i} \neq \mathcal{B}_{A_i}'$, then $q_{A_i}''=q_{A_i}$ with probability $1/2$.
			\end{itemize}
			Thus the winning probability of $\mathcal{A}$ is 
			\begin{equation*} \label{eq-mitm-3party}
			\begin{split}
			&\Pr(q_{A_i}''=q_{A_i}) \\
			& = \Pr(q_{A_i}''=q_{A_i}|~\mathcal{B}_{A_i} = \mathcal{B}_{A_i}')\Pr(\mathcal{B}_{A_i} = \mathcal{B}_{A_i}') + \Pr(q_{A_i}''=q_{A_i}|~\mathcal{B}_{A_i} \neq  \mathcal{B}_{A_i}') \Pr(\mathcal{B}_{A_i} \neq \mathcal{B}_{A_i}')\\
			&= \frac{1}{2}\{\Pr(q_{A_i}''=q_{A_i}|~\mathcal{B}_{A_i} = \mathcal{B}_{A_i}') + \Pr(q_{A_i}''=q_{A_i}|~\mathcal{B}_{A_i} \neq  \mathcal{B}_{A_i}')\}\\
			&=  \frac{1}{2}[\Pr(q_{A_i}''=q_{A_i}|~\mathcal{B} = \mathcal{B}',~q_{A_i}=q_{A_i}') \Pr(q_{A_i}=q_{A_i}') + \\
			&~~~~~~~~~~~~~~ \Pr(q_{A_i}''=q_{A_i}|~\mathcal{B} = \mathcal{B}',~q_{A_i} \neq q_{A_i}') \Pr(q_{A_i} \neq q_{A_i}') +1/2]\\
			& = \frac{1}{2}\left[1 \times \frac{1}{2} + 0 \times \frac{1}{2} + \frac{1}{2}\right]=\frac{1}{2}.
			\end{split}
			\end{equation*}
			We can calculate the winning probabilities for $q_{B_i}$ and $q_{C_i}$ in a similar way. Hence Alice, Bob and Charlie can detect this eavesdropping with probability $1-\left(\frac{1}{2} \right)^{3\delta m}>0$.
			Again, if $\mathcal{A}$ tries to eavesdrop in the second phase of transmission of qubits (Step~\ref{msg_info} of Algorithm~\ref{3_Party_msg_recons}),  Alice, Bob and Charlie can detect it in the error estimation phase (Step~\ref{3party_error3} of Algorithm~\ref{3_Party_msg_recons}) and abort the protocol.
	\end{enumerate}
	Hence our protocol is secure against a dishonest UFP , intercept-and-resend attack, entangle-and-measure attack, DoS attack and man-in-the-middle attack.
	
	\section{Multi-Party Quantum Conference} \label{sec4}
	In this section, we generalize our three-party quantum conference protocol to a multi-party quantum conference protocol. Suppose there are $N$ ($\geqslant 3$) parties $\mathcal{P}_1, \mathcal{P}_2, \ldots, \mathcal{P}_N$; each of them wants to send one's message to the other $N-1$ parties by taking help from an untrusted $(N+1)$-th party $\mathcal{P}_{(N+1)}$, who may be an eavesdropper. Let the $m$-bit messages of $\mathcal{P}_1, \mathcal{P}_2, \ldots, \mathcal{P}_N$ be ${M_1}=M_{1,1} M_{1,2}\ldots M_{1,m};\:~ {M_2}=M_{2,1} M_{2,2}\ldots M_{2,m} ;\: \ldots; \: M_N=M_{N,1} M_{N,2}\ldots M_{N,m}$ respectively, where $M_{i,j}$ is the $j$-th message bit of the $i$-th party $\mathcal{P}_i$. To do this task, first, they have to share an $m$-bit key $k=k_1k_2\ldots k_m$ and according to the key, they prepare their sequence of qubits to encode their message bits. The encoding algorithm is the same as the three-party case, i.e., Subroutine 1. Then they send their qubit sequences to $\mathcal{P}_{(N+1)}$, who measures each $N$-qubit states in $\mathcal{B}_N$ basis and announces the result publicly. Depending on the measurement results, one's message bits and key bits, each of them prepares another sequence of qubits, which contains some encoded message bits and some decoy photons, and sends it to the next party circularly. By measuring these qubits on appropriate bases, each of them gets the message bits of the previous party, but the states of the qubits corresponding to the message bits remain the same. Each adds some decoy photons to the message qubits sequence of the previous party and send it to their next party circularly and repeat this process for $N-2$ times. From the previous measurement results announced by $\mathcal{P}_{(N+1)}$, each can get other $N-1$ messages from the other $N-1$ parties. Details are given in Section~\ref{N-conf}. Note that for $N=3$, the protocol is given in Section~\ref{N-conf} reduces to the three-party protocol of Section~\ref{conf}.
	
	\subsection{Protocol 2: $N$-Party Quantum Conference}
	\label{N-conf}
	The steps of the protocol are as follows:
	\begin{enumerate}
		\item $\mathcal{P}_1, \mathcal{P}_2, \ldots, \mathcal{P}_N$ perform a Multi-party QKD protocol (e.g.,~\cite{liu2013multiparty})
		to establish an $m$ bit secret key $k=k_1k_2\ldots k_m$ between themselves.
		\item Let the $m$-bit message of $\mathcal{P}_i$ be ${M_i}=M_{i,1} M_{i,2}\ldots M_{i,m}$ for $i=1,2,\ldots,N$.
		\item For $i=1,2,\ldots,N$, the $i$-th party $\mathcal{P}_i$ prepares the sequence of qubits ${Q_i}=\{Q_i[j]\}_{j=1}^m=(Q_{i,1},Q_{i,2},\ldots ,$ $Q_{i,m})$ at its end by using the  Subroutine 1. 
		
		\item $\mathcal{P}_i$ chooses some random permutation and applies on its respective sequence of qubits $Q_i$ and get new sequence of qubits $q_i$, for $i=1,2,\ldots,N$.
		\item They send the prepared qubits $q_1,q_2,\ldots, q_N$ to $\mathcal{P}_{(N+1)}$.
		
		\item $\mathcal{P}_1, \mathcal{P}_2, \ldots, \mathcal{P}_N$ randomly choose $\delta m$ number of common positions on the sequences $Q_1 ,Q_2,$ $ \ldots, Q_N$ to estimate the error in the channel, where $\delta \ll 1$ is a small fraction. Corresponding to these rounds, they do the followings: \label{N-party_error1}
		\begin{enumerate}
			\item Each participant tells the positions and the preparation bases of those qubits for those rounds to $\mathcal{P}_{(N+1)}$. 
			\item $\mathcal{P}_{(N+1)}$ measures each single qubit states in proper bases and announces the results.
			\item $\mathcal{P}_1, \mathcal{P}_2, \ldots, \mathcal{P}_N$ reveal their respective qubits for these rounds and compare with the results announced by $\mathcal{P}_{(N+1)}$.
			\item If the estimated error is greater than some predefined threshold value, then they abort. Else they continue and go to the next step.
		\end{enumerate}
		\item $\mathcal{P}_{(N+1)}$ asks $\mathcal{P}_1, \mathcal{P}_2, \ldots, \mathcal{P}_N$ to tell the permutations which they have applied to their sequences.
		\item $\mathcal{P}_{(N+1)}$ applies the inverse permutations, corresponding to the permutations chosen by $\mathcal{P}_1, \mathcal{P}_2,$ $ \ldots, \mathcal{P}_N$, on $q_1,q_2,\ldots, q_N$ to get $Q_1 ,Q_2,\ldots, Q_N$ respectively.
		
		\item They discard the qubits corresponding to the above $\delta m$ positions. Their remaining sequences of prepared qubits are relabeled as ${Q_1}=\{Q_1[i]\}_{i=1}^{m'}$, ${Q_2}=\{Q_2[i]\}_{i=1}^{m'}$, $\ldots $, ${Q_N}=\{Q_N[i]\}_{i=1}^{m'}$, where $m'=(1-\delta) m$. 
		
		\item They update their $m$-bit key to an $m'$-bit key by discarding $\delta m$ number of key bits corresponding to the above $\delta m$ rounds. The updated key is relabeled as $k=k_1k_2\ldots k_{m'}$.
		
		\item  For $1\leqslant i \leqslant m'$, $\mathcal{P}_{(N+1)}$ measures each $N$ qubit states $Q_{1,i},Q_{2,i},\ldots ,Q_{N,i}$ in basis $\mathcal{B}_N$ and announces the result.
		
		\item $\mathcal{P}_1, \mathcal{P}_2, \ldots, \mathcal{P}_N$ make a finite sequence $\{\mathcal{M}[i]\}_{i=1}^{m'}$ containing the measurement results, i.e., for $1\leqslant i \leqslant m'$, $\mathcal{M}[i]\in \{\ket{\Phi_{0}^{+}},\ket{\Phi_{0}^{-}},\ket{\Phi_{1}^{+}},\ket{\Phi_{1}^{-}},\ldots , \ket{\Phi_{2^{(N-1)}-1}^{+}}, \ket{\Phi_{2^{(N-1)}-1}^{-}}\}$ is the $i$-th measurement result announced by $\mathcal{P}_{(N+1)}$.
		
		\item They randomly choose $\gamma m'$ number of measurement results $\mathcal{M}[i]$ from the sequence $\{\mathcal{M}[i]\}_{i=1}^{m'}$ to estimate the error, where $\gamma \ll 1$ is a small fraction.
		\begin{enumerate}
			\item They reveal their respective message bits for these rounds.
			
			\item If the estimated error is greater than some predefined threshold value, then they abort. Else they continue and go to the next step.
		\end{enumerate}
		
		\item Their remaining sequence of measurement results is relabeled as $\{\mathcal{M}[i]\}_{i=1}^{n}$, where $n=(1-\gamma) m'$.
		
		\item They update their $m'$-bit key to an $n$-bit key by discarding $\gamma m'$ number of key bits corresponding to the above $\gamma m'$ rounds. The updated key is relabeled as $k=k_1k_2\ldots k_{n}$.
		
		\item For $1\leqslant \alpha \leqslant N$, $\mathcal{P}_\alpha$ uses the Algorithm~\ref{N_Message Reconstruction} to recover others' messages.
		
	\end{enumerate} 
	
	Note that in this protocol, there are two error estimation phases. The first one checks if there is any adversary (other than $\mathcal{P}_{(N+1)}$) in the channel, who tries to get some information about the messages or change the messages. In this case, if the 1st error estimation phase does not pass, then the participants abort the protocol. Thus in this step, the motivation of $\mathcal{P}_{(N+1)}$ being correct is, there is no information gain if the parties abort the protocol. The next error estimation phase is to check, if there is any error introduced by $\mathcal{P}_{(N+1)}$.

	\subsection{Correctness and Security Analysis of $N$-Party Quantum Conference Protocol}
	In our proposed protocol, for $1\leqslant \alpha \leqslant N$, each $\mathcal{P}_\alpha$ first prepares qubits corresponding to his (her) message and shared key and then send those qubits to $\mathcal{P}_{(N+1)}$. After that, $\mathcal{P}_{(N+1)}$ measures each $N$-qubit state (one from each $\mathcal{P}_\alpha$) in basis $\mathcal{B}_N=\{\ket{\Phi_{0}^{+}}, \ket{\Phi_{0}^{-}},\ket{\Phi_{1}^{+}},\ket{\Phi_{1}^{-}}, \ldots,$  $  \ket{\Phi_{2^{(N-1)}-1}^{+}},\ket{\Phi_{2^{(N-1)}-1}^{-}}\}$  and announces the result.
	
	Now for $1 \leqslant i \leqslant m$, if $k_i=0$ (i.e preparation basis of each ${Q^\alpha}_i$ is $\{\ket{0},\ket{1}\}$) and the $N$-qubit state is $\ket{j}=\ket{j_1}\ket{j_2}\ldots \ket{j_N}$ or $\ket{2^N-1-j}=\ket{j'}=\ket{{j'}_1}\ket{{j'}_2}\ldots $  $\ket{{j'}_N}$, then after measurement, $\mathcal{P}_{(N+1)}$ will get $\ket{\Phi_{j}^{+}}$ and $\ket{\Phi_{j}^{-}} $ with probability $1/2$.
	
	Again if $k_i=1$ (i.e., the preparation basis of each ${Q^\alpha}_i$ is $\{\ket{+},\ket{-}\}$) and there are even number of $\alpha$, such that $Q_{\alpha,i}=\ket{-}$, then $\mathcal{P}_{(N+1)}$ will get $\ket{\Phi_{j}^{+}}$ ($j \in \{0,1,\ldots ,2^{(N-1)}-1\}$) with probability $1/{2^{(N-1)}}$. 
	
	Else if $k_i=1$ (i.e., preparation basis of each ${Q^\alpha}_i$ is $\{\ket{+},\ket{-}\}$) and there are odd number of $\alpha$, such that $Q_{\alpha,i}=\ket{-}$, then $\mathcal{P}_{(N+1)}$ will get $\ket{\Phi_{j}^{-}}$ ($j \in \{0,1,\ldots ,2^{(N-1)}-1\}$) with probability $1/{2^{(N-1)}}$. 
	
	For better understanding, we write the table for $N=4$ (Table~\ref{4_Party_table} in Appendix A).
	
	Now for $1 \leqslant i \leqslant m$ and $1 \leqslant \alpha \leqslant N$, if $k_i=0$, we can say the following:
	if the prepared qubit of $\mathcal{P}_\alpha$ is $\ket{0}$ or $\ket{1}$, then $\mathcal{P}_\alpha$ guesses message bit of other parties with probability $1$ as follows:
	$\mathcal{M}[i]=\ket{\Phi_{j}^{+}} \text{or } \ket{\Phi_{j}^{-}} \Rightarrow $ the $N$-qubit state was $\ket{j}$ or $\ket{2^N-1-j}$. Since $\ket{2^N-1-j}=\ket{\bar{j_1}}\ket{\bar{j_2}}\ldots \ket{\bar{j_N}}$, from his/her own message bit, $\mathcal{P}_\alpha$ can get the others' message bits.
	
	If the prepared qubit of $\mathcal{P}_\alpha$ is $\ket{+}$ or $\ket{-}$, then $\mathcal{P}_\alpha$ guesses the XOR function of message bits of all parties with probability $1$ as follows:
	\begin{equation*}
	\text{Measurement result} =
	\begin{cases}
	\ket{\Phi_{j}^{+}} \Rightarrow & M_{1,i}\oplus M_{2,i} \oplus \ldots \oplus M_{N,i}=0;\\
	\ket{\Phi_{j}^{-}} \Rightarrow & M_{1,i}\oplus M_{2,i} \oplus \ldots \oplus M_{N,i}=1.
	\end{cases}
	\end{equation*}
	for some $j \in \{0,1,\ldots ,2^{(N-1)}-1\}$.
	
	In this case, $\mathcal{P}_1,\mathcal{P}_2,\ldots,\mathcal{P}_{(\alpha-1)},\mathcal{P}_{(\alpha+2)},\ldots, \mathcal{P}_{(N-1)},\mathcal{P}_{N}$ send their encoded qubits to $\mathcal{P}_\alpha$ (encoding algorithm is given in Step~\ref{N-2nd_encoding} of Algorithm~\ref{N_Message Reconstruction}). Since $\mathcal{P}_\alpha$ knows the basis of the received qubits, by measuring the qubits in the proper basis, $\mathcal{P}_\alpha$ can know the message bits $M_{1,i}, M_{2,i},\ldots , M_{{(\alpha-1)},i}, M_{{(\alpha+2)},i},$ $\ldots, M_{N,i}$. Then from the XOR value, $\mathcal{P}_\alpha$ can get $M_{{(\alpha+1)},i}$ also.
	
	From the above discussion, we see that for all cases, all parties can conclude the communicated bits of the other parties with probability $1$. Hence our protocol is giving the correct result.
	
	The security analysis is the same as the three-party quantum conference protocol and so we will not repeat it here. 
	
		\begin{algorithm}
		\setlength{\textfloatsep}{0.05cm}
		\setlength{\floatsep}{0.05cm}
		\KwIn{Own message ${M_\alpha}$, key $k$, joint measurement results $\{\mathcal{M}[i]\}_{i=1}^{n}$ announced by $\mathcal{P}_{(N+1)}$.}
		\KwOut{Others' messages $M_1, M_2,\ldots, M_{(\alpha-1)},M_{(\alpha+1)}, \ldots,M_N$.}
		\begin{small}
			\begin{enumerate}
				\item For $1\leqslant i \leqslant n$, if $k_i = 0$,\\ 
				$\mathcal{P}_\alpha$ can learn the $i$-th bit of others' messages from the measurement result $\mathcal{M}[i]$ and his(her) own message (same as three party quantum conference, e.g., see Table~\ref{4_Party_table} for $N=4$).
				\item For $1\leqslant i \leqslant n$, if  $k_i = 1$,\\
				from the measurement result $\mathcal{M}[i]$ and his (her) own message, $\mathcal{P}_\alpha$ can learn the XOR value of the $i$-th bit of all $N$ messages. If $\mathcal{M}[i]=\ket{\Phi_{l}^{+}}$ for some $l \in \{0,1,\ldots,2^{(N-1)}-1\}$, then the value of $\chi_i=M_{1,i} \oplus M_{2,i} \oplus \ldots \oplus M_{N,i}$ becomes $0$, else $\chi_i=1$. Let $c=wt(k)$.
				\begin{enumerate}
					\item $\mathcal{P}_\alpha$ prepares an ordered set of $c$ qubits $S_{\alpha}$, corresponding to his (her) message bit where the key bit is $1$. He (she) prepares the qubits at his (her) end according to the following strategy. For $1\leqslant j \leqslant c$ and if $k_i=1$ is the $j$-th $1$ in $k$, then
					\begin{itemize} \label{N-2nd_encoding}
						\item if $M_{\alpha,i} = 0$ and $i$ is even, prepares $S_{\alpha}[j]=\ket{0}$.
						\item if $M_{\alpha,i} = 1$ and $i$ is even, prepares $S_{\alpha}[j]=\ket{1}$.
						\item if $M_{\alpha,i} = 0$ and $i$ is odd, prepares $S_{\alpha}[j]=\ket{+}$.
						\item if $M_{\alpha,i} = 1$ and $i$ is odd, prepares $S_{\alpha}[j]=\ket{-}$.
					\end{itemize}
					\item There are $N-2$ rounds.
					\begin{itemize}
						\item \textbf{$1$st round:}
						\begin{enumerate}[label={1-\arabic*.}]
							\item $\mathcal{P}_\alpha$ prepares a set of decoy photons $D_{\alpha,1}$, where the decoy photons are randomly chosen from $\{\ket{0},\ket{1},\ket{+},\ket{-}\}$. He (she) randomly inserts his (her) decoy photons into $S_{\alpha}$ and makes new ordered sets ${S_{\alpha}}^1$. $\mathcal{P}_\alpha$ sends ${S_{\alpha}}^1$ to $\mathcal{P}_{(\alpha+1)  (Mod ~N)}$ and receives ${S^1_{(\alpha-1)(Mod ~N)}}$ from $\mathcal{P}_{(\alpha-1)(Mod ~N)}$.
							\item After $\mathcal{P}_{(\alpha+1) (Mod ~N)}$ receives ${S_{\alpha}}^1$, $\mathcal{P}_\alpha$ sends the positions and states of $D_{\alpha,1}$ to $\mathcal{P}_{(\alpha+1)  (Mod ~N)}$ through a public channel. Also $\mathcal{P}_\alpha$ receives the positions and states of $D_{(\alpha-1) (Mod ~N),1}$.
							\item Then $\mathcal{P}_\alpha$ verifies the decoy photons to check eavesdropping. If there exists any eavesdropper in the quantum channel it aborts the protocol, else it goes to the next step.
							\item $\mathcal{P}_\alpha$ measures the qubits of $S_{(\alpha-1) (Mod ~N)}$ in proper bases and knows the corresponding message bits of $\mathcal{P}_{(\alpha-1) (Mod ~N)}$. Also after measurements in the proper bases, the states of the qubits of $S_{(\alpha-1) (Mod ~N)}$ remain unchanged.
						\end{enumerate}
						\item \textbf{$l$-th round ($2\leqslant l \leqslant N-2$):}
						\begin{enumerate}[label={l-\arabic*.}]
							\item $\mathcal{P}_\alpha$ prepares a set of decoy photons $D_{\alpha,l}$, where the decoy photons are randomly chosen from $\{\ket{0},\ket{1},\ket{+},\ket{-}\}$. He (she) randomly inserts his (her) decoy photons into $S_{(\alpha-l+1) (Mod ~N)}$ and makes new ordered sets ${S_{\alpha}}^l$. $\mathcal{P}_\alpha$ sends ${S_{\alpha}}^l$ to $\mathcal{P}_{(\alpha+1)  (Mod ~N)}$ and receives ${S_{(\alpha-1) (Mod ~N)}}^l$ from $\mathcal{P}_{(\alpha-1) (Mod ~N)}$.
							\item After $\mathcal{P}_{(\alpha+1)  (Mod ~N)}$ receives ${S_{\alpha}}^l$, $\mathcal{P}_\alpha$ sends the positions and states of $D_{\alpha,l}$ to $\mathcal{P}_{(\alpha+1)  (Mod ~N)}$ through a public channel. Also $\mathcal{P}_\alpha$ receives the positions and states of $D_{(\alpha-1) (Mod ~N),l}$.
							\item Then $\mathcal{P}_\alpha$ verifies the decoy photons to check eavesdropping. If there exists any eavesdropper in the quantum channel, it aborts the protocol. Else it goes to the next step.
							\item $\mathcal{P}_\alpha$ measures the qubits of $S_{(\alpha-l+1) (Mod ~N)}$ in proper bases and knows the corresponding message bits of $\mathcal{P}_{(\alpha-l+1) (Mod ~N)}$. Also after measurements in the proper bases, the states of the qubits of $S_{(\alpha-l+1) (Mod ~N)}$ remain unchanged.
						\end{enumerate}
					\end{itemize}
					\item $\mathcal{P}_\alpha$ gets all the message bits of previous $N-2$ participants. As $\mathcal{P}_\alpha$ knows $\chi_i$ and its own message bit, it gets all the other $N-1$ message bits.
				\end{enumerate}
			\end{enumerate}
			
		\end{small}
		\caption{$N$-Party Message Reconstruction Algorithm for $\mathcal{P}_\alpha$. }
		\label{N_Message Reconstruction}
		\setlength{\textfloatsep}{0.05cm}
		\setlength{\floatsep}{0.05cm}
	\end{algorithm}

	\section{Multi-party XOR Computation} \label{sec5}
	In this section, we present a protocol for multi-party XOR computation. Suppose there are $N$ parties $\mathcal{P}_1, \mathcal{P}_2, \ldots, \mathcal{P}_N$; each of them has an $m$-bit number. Let $m$-bit numbers of $\mathcal{P}_1, \mathcal{P}_2, \ldots, \mathcal{P}_N$ be ${M_1}=M_{1,1} M_{1,2}\ldots M_{1,m};\:~ {M_2}=M_{2,1} M_{2,2}\ldots M_{2,m} ;\: \ldots; \: M_N=M_{N,1} M_{N,2}\ldots $ $ M_{N,m}$ respectively, where $M_{i,j}$ is the $j$-th bit of the $i$-th party $\mathcal{P}_i$'s message. They want to compute $M_1\oplus M_2\oplus \ldots \oplus M_N$ securely, such that their numbers remain private. To execute this protocol, they will take help from an untrusted $(N+1)$-th party (or $\mathcal{P}_{(N+1)}$). Also, one participant among $\mathcal{P}_1, \mathcal{P}_2, \ldots, \mathcal{P}_N$, must be semi-honest (i.e., it follows the protocol properly), who have to play a vital role in this computation. Let $\mathcal{P}_1$ be the semi-honest participant. Other participants are only allowed to prepare and send the states corresponding to their numbers. If other participants do not follow the protocol properly (i.e., they will prepare states corresponding to a number other than their own numbers), then the computed value will be incorrect, which they definitely do not want. 
	
	To compute $M_1\oplus M_2\oplus \ldots \oplus M_N$, first $\mathcal{P}_1, \mathcal{P}_2, \ldots, \mathcal{P}_N$ have to share an $2m$-bit key $k=k_1k_2\ldots k_{2m}$ and according to the key they prepare their sequence of qubits to encode their numbers. The encoding algorithm is almost similar to conference cases. Then they send their qubit sequences to $\mathcal{P}_{(N+1)}$, who measures each $N$-qubit states in $\mathcal{B}_N$ basis and announces the result publicly. Then from this announcement and the key, they get the XOR value of their numbers. Details of this protocol are given in Section~\ref{algo:XOR}.
	
	\subsection{Protocol 3: Multi-party XOR Computation}\label{algo:XOR}
	
	\textbf{Input:} The $m$-bit numbers ${M_1}=M_{1,1} M_{1,2}\ldots M_{1,m};\: {M_2}=M_{2,1} M_{2,2}\ldots M_{2,m} ;\: \ldots; \: M_N=M_{N,1}$ $ M_{N,2}\ldots M_{N,m}$ of $N$ parties $\mathcal{P}_1, \mathcal{P}_2, \ldots, \mathcal{P}_N$  respectively.
	\\
	\textbf{Output:} $M_1\oplus M_2\oplus \ldots \oplus M_N$.
	
	The steps of the protocol are as follows:
	\begin{enumerate}
		
		\item $\mathcal{P}_1, \mathcal{P}_2, \ldots, \mathcal{P}_N$ perform a Multi-party QKD protocol~\cite{matsumoto2007multiparty} to establish an $2m$ bit secret key $k=k_1k_2\ldots k_{2m}$ between themselves.
		\item 
		\begin{enumerate}
			\item If $wt(k) = m$, then calculate $c=\oplus k_i$, $1\leq i \leq 2m$.
			\item Else if  $wt(k)>m$, then $c=1$. 
			\item Else $c=0$.
		\end{enumerate}
		\item $\mathcal{P}_1$ prepares an $m$-bit random number $k'=k'_1k'_2 \ldots k'_m$ and sends it to $\mathcal{P}_2, \ldots, \mathcal{P}_N$ by using Algorithm~\ref{send_number} with the inputs $k'$ and $k$.
		\item $\mathcal{P}_1$ calculates $M_{1_\Delta}=M_1 \oplus k'$ and uses $M_{1_\Delta}$ as his/her number.
		\item $\mathcal{P}_1$ generates a $2m$ bit string ${M'}_1$ from his/her number and the key in such a way that, for $1\leq i \leq 2m$ and $1\leq j \leq m$:
		\begin{enumerate}
			\item if $k_i=c$ and $j < m$, then ${M'}_{1,i}=M_{1_\Delta,j}$, $i=i+1$, $j=j+1$;
			\item else, ${M'}_{1,i}=x$, where $x \in \{0,1\}$ is random and $i=i+1$.
		\end{enumerate}
		
		\item For $2 \leqslant \alpha \leqslant N$: $\mathcal{P}_\alpha$ generates $2m$ bit string ${M'}_\alpha$ from his/her own number as follows. For $1\leq i \leq 2m$ and $1\leq j \leq m$: 
		\begin{enumerate}
			\item if $k_i=c$ and $j < m$, then ${M'}_{\alpha,i}=M_{\alpha,j}$, $i=i+1$, $j=j+1$;
			\item else, ${M'}_{\alpha,i}=x$, where $x \in \{0,1\}$ is random and $i=i+1$.
		\end{enumerate}
		
		\item Each $\mathcal{P}_1, \mathcal{P}_2, \ldots, \mathcal{P}_N$ prepares the sequence of qubits ${Q_1}=\{Q_1[i]\}_{i=1}^{2m}=(Q_{1,1},Q_{1,2},\ldots , $ $Q_{1,{2m}});$ ${Q_2}=\{{Q_2}[i]\}_{i=1}^{2m}=(Q_{2,1},Q_{2,2},\ldots, Q_{2,{2m}});\: \ldots ;\: {Q_N}=\{{Q_N}[i]\}_{i=1}^{2m}=(Q_{N,1},Q_{N,2},$ $ \ldots, Q_{N,{2m}})$ at their end by using Algorithm~\ref{xor_encode}.
		
		\item $\mathcal{P}_1, \mathcal{P}_2, \ldots, \mathcal{P}_N$ choose some random permutations and apply those on their respective sequences of qubits $Q_1 ,Q_2,\ldots, Q_N$ and get new sequences of qubits $q_1,q_2,\ldots, q_N$. They send their prepared sequences of qubits $q_1,q_2,\ldots, q_N$ to $\mathcal{P}_{(N+1)}$.
		
		\item $\mathcal{P}_1, \mathcal{P}_2, \ldots, \mathcal{P}_N$ randomly choose $2\delta m$ number of common positions on sequences $Q_1 ,Q_2, \ldots,$ $ Q_N$ to estimate the error in the channel, where $\delta \ll 1$ is a small fraction. Corresponding to these rounds, they do the followings: \label{XOR_error1}
		\begin{enumerate}
			\item Each participant tells the positions and preparation bases of those qubits for those rounds to $\mathcal{P}_{(N+1)}$. 
			\item $\mathcal{P}_{(N+1)}$ measures each single qubit states in proper bases and announces the results.
			\item $\mathcal{P}_1, \mathcal{P}_2, \ldots, \mathcal{P}_N$ reveal their respective qubits for these rounds and compare with the results announced by $\mathcal{P}_{(N+1)}$.
			\item If the estimated error is greater than some predefined threshold value, then they abort. Else they continue and go to the next step.
		\end{enumerate}
		\item $\mathcal{P}_{(N+1)}$ asks $\mathcal{P}_1, \mathcal{P}_2, \ldots, \mathcal{P}_N$ to tell the permutations which they have applied to their sequences.
		\item $\mathcal{P}_{(N+1)}$ applies the inverse permutations, corresponding to the permutations chosen by $\mathcal{P}_1, \mathcal{P}_2,$ $ \ldots, \mathcal{P}_N$, on $q_1,q_2,\ldots, q_N$ to get $Q_1 ,Q_2,\ldots, Q_N$ respectively.
		
		\item They discard the qubits corresponding to the above $2\delta m$ positions. Their remaining sequences of prepared qubits are relabeled as ${Q_1}=\{Q_1[i]\}_{i=1}^{2m'}$, ${Q_2}=\{Q_2[i]\}_{i=1}^{2m'}$, $\ldots $, ${Q_N}=\{Q_N[i]\}_{i=1}^{2 m'}$ where $m'=(1-\delta) m$.
		
		\item They update their $2m$-bit key to an $2m'$-bit key by discarding $2\delta m$ number of key bits corresponding to the above $2\delta m$ rounds. The updated key is relabeled as $k=k_1k_2\ldots k_{2m'}$.
		
		\item  For $1\leqslant i \leqslant 2m'$, $\mathcal{P}_{(N+1)}$ measures each $N$ qubit states $Q_{1,i},Q_{2,i},\ldots ,Q_{N,i}$ in basis $\mathcal{B}_N$ and announces the result.
		
		\item $\mathcal{P}_1, \mathcal{P}_2, \ldots, \mathcal{P}_N$ make a finite sequence $\{\mathcal{M}[i]\}_{i=1}^{2m'}$ containing the measurement results, i.e., for $1\leqslant i \leqslant 2m'$, $\mathcal{M}[i]\in \{\ket{\Phi_{0}^{+}},\ket{\Phi_{0}^{-}},\ket{\Phi_{1}^{+}},\ket{\Phi_{1}^{-}},\ldots , \ket{\Phi_{2^{(N-1)}-1}^{+}}, \ket{\Phi_{2^{(N-1)}-1}^{-}}\}$ is the $i$-th measurement result announced by $\mathcal{P}_{(N+1)}$.
		
		\item They randomly choose $2\gamma m'$ number of measurement results $\mathcal{M}[i]$ from the sequence $\{\mathcal{M}[i]\}_{i=1}^{2m'}$ to estimate the error, where $\gamma \ll 1$ is a small fraction.
		\begin{enumerate}
			\item For these rounds, they reveal respective bits of their numbers.
			
			\item If the estimated error is greater than some predefined threshold value, then they abort. Else they continue and go to the next step.
		\end{enumerate}
		
		\item Their remaining sequence of measurement results is relabeled as $\{\mathcal{M}[i]\}_{i=1}^{2n}$, where $n=(1-\gamma) m'$.
		
		\item They update their $2m'$-bit key to an $2n$-bit key by discarding $2\gamma m'$ number of key bits corresponding to the above $2\gamma m'$ rounds. The updated key is relabeled as $k=k_1k_2\ldots k_{2n}$.
		\item For $1\leqslant i \leqslant 2n$,
		\begin{enumerate}
			\item if  $k_i = \bar{c}$, then each participant can learn $i$-th bit of others' number from the measurement result $\mathcal{M}[i]$ and their own number (see Algorithm~\ref{N-conf}).
			\item Else, from the measurement result $\mathcal{M}[i]$, each participant can learn the XOR value of the $i$-th bit of all $N$ numbers.  If $\mathcal{M}[i]=\ket{\Phi_{l}^{+}}$ for some $l \in \{0,1,\ldots,2^{(N-1)}-1\}$, then the value of $\chi_i=M_{1_\Delta,i} \oplus M_{2,i} \oplus \ldots \oplus M_{N,i}$ becomes $0$, else $\chi_i=1$.\label{xor__msg_info}
		\end{enumerate} 
		\item Combining the knowledges from Step-\ref{xor__msg_info} and the key, they can get $M_{1_\Delta} \oplus M_2 \oplus \ldots \oplus M_N$. 
		\item $\mathcal{P}_1, \mathcal{P}_2, \ldots, \mathcal{P}_N$ calculate $M_1 \oplus M_2 \oplus \ldots \oplus M_N= k' \oplus  M_{1_\Delta} \oplus M_2 \oplus \ldots \oplus M_N$.
		
	\end{enumerate}

	\begin{algorithm}
		\setlength{\textfloatsep}{0.05cm}
		\setlength{\floatsep}{0.05cm}
		\KwIn{Random number $k'=k'_1k'_2\ldots k'_m$ chosen by $\mathcal{P}_1$, key $k=k_1k_2\ldots k_{2m}$. }
		\KwOut{For $2 \leqslant \alpha \leqslant N$, $\mathcal{P}_\alpha$ has $k'$.}
		\begin{enumerate}
			\item To encode random number $k'$, $\mathcal{P}_1$ prepares $N-1$ sets of qubits $Q_\alpha=Q_{\alpha,1}Q_{\alpha,2}\ldots Q_{\alpha,m}$  for $\mathcal{P}_\alpha$ ($2 \leqslant \alpha \leqslant N$), by using the following strategy: for $1 \leqslant i \leqslant m$ and $2 \leqslant \alpha \leqslant N$,
			\begin{enumerate}
				\item if $k'_i=0$ and $k_i=0 \Rightarrow Q_{\alpha,i}=\ket{0}$
				\item if $k'_i=1$ and $k_i=0 \Rightarrow Q_{\alpha,i}=\ket{1}$
				\item if $k'_i=0$ and $k_i=1 \Rightarrow Q_{\alpha,i}=\ket{+}$
				\item if $k'_i=1$ and $k_i=1 \Rightarrow Q_{\alpha,i}=\ket{-}$
			\end{enumerate}
			\item For $2 \leqslant \alpha \leqslant N$, $\mathcal{P}_1$ chooses a set of decoy photons $D_\alpha$ and randomly inserts those decoy photons into $Q_\alpha$ and gets new set of qubits $q_\alpha$. 
			
			\item $\mathcal{P}_1$ sends $q_\alpha$ to $\mathcal{P}_\alpha$.
			
			\item All $\mathcal{P}_\alpha$ inform $\mathcal{P}_1$ that they receive $q_\alpha$.
			
			\item $\mathcal{P}_1$ announces the positions and states of the decoy photons.
			
			\item Each $\mathcal{P}_\alpha$ measures the decoy photons in their appropriate bases and calculate the error in the channel (or check that if there is any eavesdropper). 
			
			\item If the error rate is in a tolerable range, then $\mathcal{P}_\alpha$ measures the qubits of $Q_\alpha$ in their appropriate bases (determined by the key) and get $k'$.
			
		\end{enumerate}
		\caption{Algorithm for Sending a Number to $(N-1)$-Participant.}
		\label{send_number}
		\setlength{\textfloatsep}{0.05cm}
		\setlength{\floatsep}{0.05cm}
	\end{algorithm}
	
	\begin{algorithm}
		\setlength{\textfloatsep}{0.05cm}
		\setlength{\floatsep}{0.05cm}
		\KwIn{$M'_\alpha$ = $2m$-bit message of $\mathcal{P}_\alpha$, key $k=k_1k_2\ldots k_{2m}$. }
		\KwOut{Sequence of qubits ${Q_\alpha}=\{Q_\alpha[i]\}_{i=1}^{2m}=(Q_{\alpha,1},Q_{\alpha,2},\ldots ,Q_{\alpha,{2m}})$.}

		\begin{enumerate}
			\item 
		\end{enumerate}
		\begin{enumerate}
			\item
			\begin{enumerate}
				\item If $wt(k) = m$, then calculate $c=\oplus k_i$, $1\leq i \leq 2m$.
				\item Else if  $wt(k)>m$, then $c=1$. 
				\item Else $c=0$.
			\end{enumerate}
			\item For $1 \leqslant i \leqslant 2m$,
			
			\begin{enumerate}
				\item if ${M'}_{\alpha,i} = 0$ and $k_i = \bar{c}$, set $Q_{1,i}$ (or $Q_{2,i} \ldots $ or $Q_{N,i}=\ket{0}$;
				\item if ${M'}_{\alpha,i}= 1$ and $k_i = \bar{c}$, set $Q_{1,i}$ (or $Q_{2,i} \ldots $ or $Q_{N,i}=\ket{1}$;
				\item if ${M'}_{\alpha,i}= 0$ and $k_i = {c}$, set $Q_{1,i}$ (or $Q_{2,i} \ldots $ or $Q_{N,i}=\ket{+}$;
				\item if ${M'}_{\alpha,i}= 1$ and $k_i = c$, set $Q_{1,i}$ (or $Q_{2,i} \ldots $ or $Q_{N,i}=\ket{-}$.
			\end{enumerate}
		\end{enumerate}
		\caption{Message Encoding Algorithm for Multi-party XOR Computation.}
		\label{xor_encode}
		\setlength{\textfloatsep}{0.05cm}
		\setlength{\floatsep}{0.05cm}
	\end{algorithm}

	\subsection{Correctness and Security Analysis of the Quantum Protocol for Multi-party XOR computation}
	The correctness of this protocol directly follows from the previous one (i.e., multi-party quantum conference protocol). Also, we can say this protocol is secure against intercept-and-resend attack, disturbance attack, entangle-and-measure attack, and dishonest $\mathcal{P}_{(N+1)}$, as this is a part of the previous protocol discussed in the last section.
	
	Now, we only have to prove that, no one can get the computed XOR-value other than the legitimate parties.
	
	Let an adversary \textit{A} constructs a $2m$-bit string $\tau=\tau_1\tau_2\ldots \tau_{2m}$, from the measurement results in such a way that, if $\mathcal{M}[i]=\ket{\Phi_{l}^{+}}$ for some $l \in \{0,1,\ldots,2^{(N-1)}-1\}$, then $\tau_i=0$, else if $\mathcal{M}[i]=\ket{\Phi_{l}^{-}}$ for some $l \in \{0,1,\ldots,2^{(N-1)}-1\}$, then $\tau_i=1$. Now $m$-bit string $\eta=M_{1_\Delta}\oplus M_2\oplus \ldots \oplus M_N$ is a subsequence of $\tau$. If \textit{A} can guess $\eta$ from $\tau$ with some low probability, then also it can not get any information about $\mu= M_1\oplus M_2\oplus \ldots \oplus M_N$ as $\mu=\eta \oplus k'$, where $k'$ is unknown to him/her. Then from the notion of security of the famous ``one time pad"  protocol~\cite{Shannon1949}, we can say that our proposed protocol is secure.
	
	It is to be noted that, if $\mathcal{P}_1$ is dishonest, then he/she can cheat and get the exact XOR value, whereas the other participants get some random value instead of the exact XOR value. This thing happens in the following way: $\mathcal{P}_1$ calculates $M_{1_\Delta}=M_1 \oplus R$, where $R\neq k'$ is a random $m$-bit number and it is used instead of $k'$. Then $\mathcal{P}_1$ follows all the next steps of the protocol. At the end of the protocol, everyone get $M_{1_\Delta}\oplus M_2 \oplus \ldots \oplus M_N$. Then $\mathcal{P}_2, \ldots, \mathcal{P}_N$ calculate $M_1 \oplus M_2 \oplus \ldots \oplus M_N= k' \oplus  M_{1_\Delta} \oplus M_2 \oplus \ldots \oplus M_N$, which is not true as $R\neq k'$. But, $\mathcal{P}_1$ calculates $M_1 \oplus M_2 \oplus \ldots \oplus M_N = R \oplus  M_{1_\Delta} \oplus M_2 \oplus \ldots \oplus M_N$, which is correct. That is, after executing the protocol, $\mathcal{P}_1$ has the exact value of $M_1 \oplus M_2 \oplus \ldots \oplus M_N$ and other participants have the value of $k' \oplus R \oplus M_1 \oplus M_2 \oplus \ldots \oplus M_N$, which is nothing but a random number.
	
	Thus here we are assuming that $\mathcal{P}_1$ is semi-honest, that is,  follows the protocol properly. Hence each participant gets the computed XOR-value exactly, but no other party can not get any information about the value.
	
	\section{Conclusion} \label{sec6}
	In this paper, first we identify that the MDI-QD protocol presented in~\cite{qip/Maitra17} is not secure against the intercept-and-resend attack, and we modify the protocol to make it secure against this attack. Then we present three more protocols, two of them for the quantum conference, i.e., securely and simultaneously exchanging secret messages between the participants. The first protocol is for three parties and then we generalize it to a multi-party scenario, i.e., for $N$-parties (where $N \geqslant 3$). Another protocol presented in this paper is for multi-party XOR computation, where $N$-parties can compute the XOR function of their own numbers, but their numbers remain private. All the protocols discussed above are proven to be correct and secure.


\begin{thebibliography}{000}

\bibitem{wootters1982single}
W.K. Wootters and  W.H. Zurek (1982), {\it A single quantum cannot be cloned}, Nature, 299(5886), pp. 802-803.
	
\bibitem{tcs/BennettB14}
C.H. Bennett and  G. Brassard (2014), {\it Quantum cryptography: public key distribution and coin tossing}, Theor. Comput. Sci., 560(12), pp. 7-11.
	
\bibitem{shor2000simple}
 P.W. Shor and  J. Preskill (2000), {\it Simple proof of security of the BB84 quantum key distribution protocol}, Physical review letters, 85(2), p. 441.

\bibitem{ekert1991quantum}
A. K. Ekert (1991), {\it Quantum cryptography based on Bell's theorem}, Physical review letters, 67(6), p. 661.
	
\bibitem{Brassard1992quantum}
C.H. Bennett, G. Brassard and N.D. Mermin (1992), {\it Quantum cryptography without Bell's theorem}, Physical review letters, 68(5), p. 557.

\bibitem{bennett1992quantum}
C.H. Bennett (1992), {\it Quantum cryptography using any two nonorthogonal states.}, Physical review letters, 68(21), p. 3121.

\bibitem{long2002theoretically}
G.L. Long and X.S. Liu (2002), {\it Theoretically efficient high-capacity quantum-key-distribution
	scheme}, Physical Review A, 65(3), p. 032302.	
	
\bibitem{xue2002conditional}
P. Xue,  C.F. Li, and G.C. Guo (2002), {\it Conditional efficient multiuser quantum cryptography network}, Physical Review A, 65(2), p. 022317.	
	
\bibitem{deng2004bidirectional}
F.G. Deng and G.L. Long (2004), {\it Bidirectional quantum key distribution protocol with practical faint laser pulses}, Physical Review A, 70(1), p. 012311.

\bibitem{hwang2003quantum}
 W.Y. Hwang (2003), {\it Quantum key distribution with high loss: toward global secure
	communication.}, Physical Review Letters, 91(5), p. 057901.

\bibitem{lo2005decoy}
 H.K. Lo, X. Ma and K. Chen (2005), {\it Decoy state quantum key distribution}, Physical review letters, 94(23), p. 230504.
	
\bibitem{lo2012measurement}
H.K. Lo, M. Curty and B. Qi (2012), {\it Measurement-device-independent quantum key distribution}, Physical review letters, 108(13), p. 130503.
	
\bibitem{barrett2005no}
J. Barrett, L. Hardy and A. Kent (2005), {\it No signaling and quantum key distribution}, Physical review letters, 95(1), p. 010503.

\bibitem{grosshans2003quantum}
F. Grosshans, G. Van Assche, J. Wenger, R. Brouri, N.J. Cerf and P. Grangier (2003), {\it Quantum key distribution using gaussian-modulated coherent states}, Nature, 421(6920), pp. 238-241.

\bibitem{bostrom2002ping}
K.J. Bostr{\"o}m and T. Felbinger (2005), {\it Ping-pong coding}, Phys. Rev. Lett., 89(quant-ph/0209040), p. 187902.

\bibitem{deng2003two}
F.G. Deng, G.L. Long and X.S. Liu (2003), {\it Two-step quantum direct communication protocol using the 	einstein-podolsky-rosen pair block}, Physical Review A, 68(4), p. 042317.

\bibitem{deng2004secure}
F.G. Deng and G.L. Long (2004), {\it Secure direct communication with a quantum one-time pad}, Physical Review A, 69(5), p. 052319.

\bibitem{wang2005quantum}
C. Wang, F.G. Deng,  Y.S. Li, X.S. Liu and G.L. Long (2005), {\it Quantum secure direct communication with high-dimension quantum superdense coding}, Physical Review A, 71(4), p. 044305.

\bibitem{wang2005multi}
C. Wang, F.G. Deng and G.L. Long (2005), {\it Multi-step quantum secure direct communication using multi-particle green--horne--zeilinger state}, Optics communications, 253(1-3), pp. 15-20.
	
\bibitem{wang2006quantum}
J. Wang, Q. Zhang and C.J. Tang (2006), {\it Quantum secure direct communication based on order rearrangement of
	single photons}, Physics Letters A, 358(4), pp. 256-258.
	
\bibitem{long2007quantum}
	G.L. Long, F.G. Deng, C. Wang, X.H. Li, K. Wen, and W.Y. Wang (2007), {\it Quantum secure direct communication and deterministic secure quantum communication}, Frontiers of Physics in China, 2(3), pp. 251-272.
	
\bibitem{xi2007quantum}
	L.X. Han, L.C. Yan, D.F. Guo, Z. Ping, L.Y. Jie, and Z.H. Yu (2007), {\it Quantum secure direct communication with quantum encryption based on pure entangled states}, Chinese Physics, 16(8), p. 2149.
	
\bibitem{das2020improving}
N. Das and G. Paul (2020), {\it Improving the Security of ``{M}easurement-Device-Independent Quantum Communication without Encryption"}, Science Bulletin, 65 (24), p. 2048, (https://doi.org/10.1016/j.scib.2020.09.015).

\bibitem{das2020cryptanalysis}
N. Das and G. Paul (2020), {\it Cryptanalysis of Quantum Secure Direct Communication Protocol with Mutual Authentication Based on Single Photons and Bell States}, arXiv preprint arXiv:2007.03710.
	
\bibitem{nguyen2004quantum}
	B.A. Nguyen (2004), {\it Quantum dialogue}, Physics Letters A, 328(1), pp. 6-10.
	
\bibitem{zhang2004deterministic}
	Z. Zhang (2004), {\it Deterministic secure direct bidirectional communication protocol}, arXiv preprint quant-ph/0403186.
	
\bibitem{zhong2005quantum}
	M.Z. Xiao, Z.Z. Jun, and L. Yong (2005), {\it Quantum dialogue revisited}, Chinese Physics Letters, 22(1), p. 22.
	
\bibitem{xia2006quantum}
	Y. Xia, C.B. Fu, S. Zhang, S.K. Hong, K.H. Yeon, and C.I. Um (2006), {\it Quantum dialogue by using the ghz state}, arXiv preprint quant-ph/0601127.
	
\bibitem{xin2006secure}
	J. Xin and Z. Shou (2006), {\it Secure quantum dialogue based on single-photon}, Chinese Physics, 15(7), p. 1418.
	
\bibitem{yan2007controlled}
	X. Yan, S. Jie, N. Jing and S.H. Shan (2007), {\it Controlled secure quantum dialogue using a pure entangled ghz states}, Communications in Theoretical Physics, 48(5), p. 841.
	
\bibitem{tan2008classical}
	Y. G. Tan and Q.Y. Cai (2008), {\it Classical correlation in quantum dialogue}, International Journal of Quantum Information, 6(02), pp. 325-329.
	
\bibitem{gao2008revisiting}
	F. Gao, F. Guo, Q. Wen and F. Zhu (2008), {\it Revisiting the security of quantum dialogue and bidirectional quantum	secure direct communication}, Science in China Series G: Physics, Mechanics and Astronomy, 51(5), pp.559-566.
		
\bibitem{gao2010two}
	G. Gao (2010), {\it Two quantum dialogue protocols without information leakage}, Optics communications, 283(10), pp. 2288-2293.
		
\bibitem{qip/Maitra17}
	A. Maitra (2017), {\it Measurement device-independent quantum dialogue}, Quantum Information Processing, 16(12), p. 305.

\bibitem{das2020two}
N. Das and G. Paul (2020), {\it Two Efficient Measurement Device Independent Quantum Dialogue Protocols}, International Journal of Quantum Information, 2050038.

\bibitem{gao2005deterministic}
	T. Gao, F.L. Yan and Z.X. Wang (2005), {\it Deterministic secure direct communication using ghz states and
		swapping quantum entanglement}, Journal of Physics A: Mathematical and General, 38(25), p. 5761.
	
\bibitem{jin2006three}
	X.R. Jin, X. Ji, Y.Q. Zhang, S. Zhang, S.K. Hong, K.H. Yeon and C.I. Um (2006), {\it Three-party quantum secure direct communication based on ghz states}, Physics Letters A, 354(1-2), pp. 67-70.
	
\bibitem{ting2005simultaneous}
	G. Ting, Y.F. Li and W.Z. Xi (2005), {\it A simultaneous quantum secure direct communication scheme between the
		central party and other m parties}, Chinese Physics Letters, 22(10), p. 2473.
	
\bibitem{tan2014multi}
	X. Tan, X. Zhang and C. Liang (2014), {\it Multi-party quantum secure direct communication}, In 2014 Ninth International Conference on P2P, Parallel, Grid, Cloud and Internet Computing (pp. 251-255). IEEE.
	
\bibitem{zhang2005multiparty}
	Z.J. Zhang, Y. Li and Z.X Man (2005), {\it Multiparty quantum secret sharing}, Physical Review A, 71(4), p. 044301.
	
\bibitem{banerjee2018quantum}
	A. Banerjee, K. Thapliyal, C. Shukla and A. Pathak (2018), {\it Quantum conference}, Quantum Information Processing, 17(7), p. 161.
	
\bibitem{hillery1999quantum}
	M. Hillery, V. Bu{\v{z}}ek and A. Berthiaume (1999), {\it Quantum secret sharing}, Physical Review A, 59(3), p. 1829.
	
\bibitem{zhang2005multiparty_qss}
	Z.J. Zhang (2005), {\it Multiparty quantum secret sharing of secure direct communication}, Physics Letters A, 342(1-2), pp. 60-66.
	
\bibitem{gottesman2000theory}
	D. Gottesman (2000), {\it Theory of quantum secret sharing}, Physical Review A, 61(4), p. 042311.
	
\bibitem{guo2003quantum}
	G.P. Guo and G.c. Guo (2003), {\it Quantum secret sharing without entanglement}, Physics Letters A, 310(4), pp. 247-251.
	
\bibitem{shi2016secure}
	R.H Shi, Y. Mu, H. Zhong, J. Cui and S. Zhang (2016), {\it Secure multiparty quantum computation for summation and multiplication}, Scientific reports, 6(1), pp. 1-9.
	
\bibitem{chen2010} X.B. Chen,  G. Xu, Y.X. Yang  and  Q.Y. Wen (2010), {\it An efficient protocol for the secure multi-party quantum summation}, International Journal of Theoretical Physics, 49(11), pp. 2793-2804.

\bibitem{Liu2013} W. Liu,  C. Liu, H Wang  and T. Jia (2013), {\it Quantum private comparison: a review} IETE Technical Review, 30(5), pp. 439-445. 

\bibitem{Zhang2013} W.W. Zhang,  and  K.J. Zhang (2013) {\it Cryptanalysis and improvement of the quantum private comparison protocol with semi-honest third party}, Quantum information processing, 12(5), pp. 1981-1990.

\bibitem{liu2015} W. Liu,  Y.B. Wang  and  X.M. Wang (2015), {\it Quantum multi-party private comparison protocol using d-dimensional Bell states}, International Journal of Theoretical Physics, 54(6), pp. 1830-1839.
	
\bibitem{matsumoto2007multiparty}
	R. Matsumoto (2007), {\it Multiparty quantum-key-distribution protocol without use of
		entanglement}, Physical Review A, 76(6), p. 062316.
	
\bibitem{liu2013multiparty}
	B. Liu, F. Gao, W. Huang and Q.Y Wen(2013), {\it Multiparty quantum key agreement with single particles}, Quantum information processing, 12(4), pp. 1797-1805.
	
\bibitem{Shannon1949}
	C.E. Shannon (1949), {\it Communication theory of secrecy systems}, Bell system technical journal, 28(4), pp. 656-715.
\bibitem{nielsen2002quantum}  M.A. Nielsen and I. Chuang (2002) {\it Quantum computation and quantum information}.

\end{thebibliography}

\appendix
\section*{Appendix}
\begin{table}[h]
	\caption{Comparison between quantum conference proposed in~\cite{banerjee2018quantum} and our protocol.}
	\begin{tabular}{|l|l|}
		\hline
		\textbf{Quantum conference~\cite{banerjee2018quantum}}                   & \textbf{Our protocol}                           \\ \hline
		Uses $2^m(N-1)$ unitary operators             & No unitary operator                             \\ \hline
		$n$-qubit entangled state, $n \geq (N-1)m$    & Single qubit states                             \\ \hline
		Approximate $1$ qubit for $1$-bit information & Approximate $3/2$ qubit for $1$-bit information \\ \hline
		No key required                               & One initial key is required                     \\ \hline
	\end{tabular}
\end{table}
\noindent
\pagestyle{empty}
	
	\begin{sidewaystable}
			\centering
			\renewcommand*{\arraystretch}{1.6}
			\caption{Different cases in Four Party Quantum Conference.}
			\setlength{\tabcolsep}{8pt}
			\resizebox{1.0\textwidth}{!}{
				\begin{tabular}{|c|c|c|c|c|c|c|c|c|c|c|c|c|c|c|c|c|c|c|c|c|c|c|c|}
				\hline
				\multicolumn{4}{|c|}{Qubits sent by} &  \multicolumn{16}{|c|}{Probability (Eve's end)} & \multicolumn{4}{|c|}{Communicated Bits}\\
				\hline
				{} $\mathcal{P}_1$ & $\mathcal{P}_2$ & $\mathcal{P}_3$ & $\mathcal{P}_4$  & $\ket{\phi_0^+}$ & $\ket{\phi_0^-}$ & $\ket{\phi_1^+}$ & $\ket{\phi_1^-}$ & $\ket{\phi_2^+}$ & $\ket{\phi_2^-}$ & $\ket{\phi_3^+}$& $\ket{\phi_3^-}$   &$\ket{\phi_4^+}$ & $\ket{\phi_4^-}$ & $\ket{\phi_5^+}$ & $\ket{\phi_5^-}$ & $\ket{\phi_6^+}$ & $\ket{\phi_6^-}$ & $\ket{\phi_7^+}$& $\ket{\phi_7^-}$   & by $\mathcal{P}_1$   & by $\mathcal{P}_2$ & by $\mathcal{P}_3$ & by $\mathcal{P}_4$\\ $\ket{0}$  & $\ket{0}$  &  $\ket{0}$ &  $\ket{0}$ & $1/2$ & $1/2$  & $0$ & $0$ & $0$ & $0$ & $0$ & $0$  & $0$  & $0$  & $0$  & $0$  & $0$  & $0$  & $0$  & $0$  & $0$ & $0$ & $0$ & $0$  \\
				$\ket{0}$  & $\ket{0}$  &  $\ket{0}$ &  $\ket{1}$ &  $0$ & $0$ & $1/2$ & $1/2$ & $0$ & $0$ & $0$  & $0$ & $0$  & $0$  & $0$  & $0$  & $0$  & $0$  & $0$  & $0$  & $0$  & $0$ & $0$ & $1$\\
				$\ket{0}$  & $\ket{0}$  &  $\ket{1}$ &  $\ket{0}$ &  $0$ & $0$ & $0$ & $0$ & $1/2$ & $1/2$ & $0$ & $0$ & $0$  & $0$  & $0$  & $0$  & $0$  & $0$  & $0$  & $0$  & $0$  & $0$  & $1$ & $0$\\
				$\ket{0}$  & $\ket{0}$  &  $\ket{1}$ &  $\ket{1}$  & $0$ & $0$ & $0$ & $0$ & $0$ & $0$ & $1/2$ & $1/2$ & $0$  & $0$  & $0$  & $0$  & $0$  & $0$  & $0$  & $0$  & $0$  & $0$ & $1$ & $1$\\
				$\ket{0}$  & $\ket{1}$  &  $\ket{0}$ &  $\ket{0}$ & $0$ & $0$ & $0$ & $0$ & $0$ & $0$ & $0$  & $0$  & $1/2$ & $1/2$ & $0$  & $0$  & $0$  & $0$  & $0$  & $0$  & $0$  & $1$ & $0$ & $0$\\
				$\ket{0}$  & $\ket{1}$  &  $\ket{0}$ &  $\ket{1}$ & $0$ & $0$ & $0$ & $0$ & $0$  & $0$  & $0$  & $0$ & $0$  & $0$ & $1/2$ & $1/2$   & $0$  & $0$  & $0$  & $0$ & $0$ & $1$  & $0$ & $1$\\
				$\ket{0}$  & $\ket{1}$  &  $\ket{1}$ &  $\ket{0}$ &  $0$& $0$  & $0$  & $0$  & $0$  & $0$  & $0$  & $0$ & $0$  & $0$  & $0$  & $0$ &  $1/2$ & $1/2$ & $0$ & $0$  & $0$ & $1$ & $1$ & $0$ \\
				$\ket{0}$  & $\ket{1}$  &  $\ket{1}$ &  $\ket{1}$ & $0$ & $0$ & $0$ & $0$ & $0$ & $0$ & $0$ & $0$ & $0$ & $0$ & $0$ & $0$ & $0$ & $0$ & $1/2$ & $1/2$  & $0$ & $1$ & $1$ & $1$  \\
				$\ket{1}$  &  $\ket{0}$  &  $\ket{0}$ &  $\ket{0}$& $0$ & $0$ & $0$ & $0$ & $0$ & $0$ & $0$ & $0$ & $0$ & $0$ & $0$ & $0$ & $0$ & $0$ & $1/2$ & $1/2$  & $1$ & $0$ & $0$ & $0$  \\
				$\ket{1}$  &  $\ket{0}$  &  $\ket{0}$ &  $\ket{1}$ &$0$ & $0$ &$0$ & $0$ &$0$ & $0$ &$0$ & $0$ &  $0$ & $0$ & $0$ & $0$ & $1/2$ & $1/2$ & $0$  & $0$ & $1$ & $0$ & $0$ & $1$\\
				$\ket{1}$  &  $\ket{0}$  &  $\ket{1}$ &  $\ket{0}$ &  $0$ & $0$ & $0$ & $0$ &$0$ & $0$ & $0$  &$0$ &$0$ & $0$ & $1/2$ & $1/2$ & $0$  &$0$ & $0$ & $0$ &$1$ & $0$  & $1$ & $0$\\
				$\ket{1}$  &  $\ket{0}$  &  $\ket{1}$ &  $\ket{1}$  & $0$ & $0$ & $0$ & $0$ & $0$ & $0$ & $0$  &$0$ & $1/2$ & $1/2$& $0$ & $0$ & $0$  &$0$& $0$ & $0$ &$1$ & $0$ & $1$ & $1$\\
				$\ket{1}$  &  $\ket{1}$  &  $\ket{0}$ &  $\ket{0}$ & $0$ & $0$ & $0$ & $0$ & $0$ & $0$ & $1/2$ & $1/2$& $0$ & $0$ & $0$  &$0$ & $0$ & $0$ & $0$  &$0$ &$1$ & $1$ & $0$ & $0$\\
				$\ket{1}$  &  $\ket{1}$  &  $\ket{0}$ &  $\ket{1}$ & $0$ & $0$ & $0$ & $0$ & $1/2$ & $1/2$ & $0$ & $0$& $0$ & $0$ & $0$  &$0$& $0$ & $0$ & $0$  & $0$ &$1$ & $1$  & $0$ & $1$\\
				$\ket{1}$  &  $\ket{1}$  &  $\ket{1}$ &  $\ket{0}$ &  $0$ & $0$ & $1/2$ & $1/2$ & $0$ & $0$ & $0$  &$0$& $0$ & $0$ & $0$  &$0$ & $0$ & $0$ & $0$  &$0$ & $1$ & $1$ & $1$ & $0$ \\
				$\ket{1}$  &  $\ket{1}$  &  $\ket{1}$ &  $\ket{1}$ & $1/2$ & $1/2$  & $0$ & $0$ & $0$ & $0$ & $0$ & $0$ & $0$ & $0$ & $0$ & $0$ & $0$ & $0$ & $0$ & $0$ &$1$ & $1$ & $1$ & $1$  \\
				\hline
				$\ket{+}$  &  $\ket{+}$  &  $\ket{+}$  &  $\ket{+}$   & $1/8$ & $0$ & $1/8$ & $0$ & $1/8$ & $0$  & $1/8$  &  $0$ & $1/8$ & $0$  & $1/8$ &  $0$  & $1/8$ & $0$& $1/8$ & $0$  & $0$  & $0$ & $0$ & $0$ \\
				$\ket{+}$  &  $\ket{+}$  &  $\ket{+}$  &  $\ket{-}$  & $0$  & $1/8$ & $0$ & $1/8$ & $0$ & $1/8$ & $0$  & $1/8$  &  $0$ & $1/8$ & $0$  & $1/8$ &  $0$  & $1/8$ & $0$& $1/8$ & $0$  & $0$ & $0$ & $1$\\
				$\ket{+}$  &  $\ket{+}$  &  $\ket{-}$  &  $\ket{+}$  & $0$  & $1/8$ & $0$ & $1/8$ & $0$ & $1/8$ & $0$  & $1/8$  &  $0$ & $1/8$ & $0$  & $1/8$ &  $0$  & $1/8$ & $0$& $1/8$ & $0$  &$0$  & $1$ & $0$\\
				$\ket{+}$  &  $\ket{+}$  &  $\ket{-}$  &  $\ket{-}$   & $1/8$ & $0$ & $1/8$ & $0$ & $1/8$ & $0$  & $1/8$  &  $0$ & $1/8$ & $0$  & $1/8$ &  $0$  & $1/8$ & $0$& $1/8$ & $0$ & $0$ & $0$ & $1$ & $1$\\
				$\ket{+}$  &  $\ket{-}$  &  $\ket{+}$  &  $\ket{+}$  & $0$  & $1/8$ & $0$ & $1/8$ & $0$ & $1/8$ & $0$  & $1/8$  &  $0$ & $1/8$ & $0$  & $1/8$ &  $0$  & $1/8$ & $0$& $1/8$ & $0$ & $1$ & $0$ & $0$\\
				$\ket{+}$  &  $\ket{-}$  &  $\ket{+}$  &  $\ket{-}$   & $1/8$ & $0$ & $1/8$ & $0$ & $1/8$ & $0$  & $1/8$  &  $0$ & $1/8$ & $0$  & $1/8$ &  $0$  & $1/8$ & $0$& $1/8$ & $0$  & $0$ &  $1$  & $0$ & $1$\\
				$\ket{+}$  &  $\ket{-}$  &  $\ket{-}$  &  $\ket{+}$  & $1/8$ & $0$ & $1/8$ & $0$ & $1/8$ & $0$  & $1/8$  &  $0$ & $1/8$ & $0$  & $1/8$ &  $0$  & $1/8$ & $0$& $1/8$ & $0$  & $0$ &  $1$ & $1$ & $0$ \\
				$\ket{+}$  &  $\ket{-}$  &  $\ket{-}$  &  $\ket{-}$   & $0$  & $1/8$ & $0$ & $1/8$ & $0$ & $1/8$ & $0$  & $1/8$  &  $0$ & $1/8$ & $0$  & $1/8$ &  $0$  & $1/8$ & $0$& $1/8$ & $0$ & $1$ & $1$ & $1$  \\
				$\ket{-}$  &  $\ket{+}$  &  $\ket{+}$  &  $\ket{+}$  & $0$  & $1/8$ & $0$ & $1/8$ & $0$ & $1/8$ & $0$  & $1/8$  &  $0$ & $1/8$ & $0$  & $1/8$ &  $0$  & $1/8$ & $0$& $1/8$ & $1$  & $0$ & $0$ & $0$ \\
				$\ket{-}$  &  $\ket{+}$  &  $\ket{+}$  &  $\ket{-}$   & $1/8$ & $0$ & $1/8$ & $0$ & $1/8$ & $0$  & $1/8$  &  $0$ & $1/8$ & $0$  & $1/8$ &  $0$  & $1/8$ & $0$& $1/8$ & $0$  & $1$ & $0$ & $0$ & $1$\\
				$\ket{-}$  &  $\ket{+}$  &  $\ket{-}$  &  $\ket{+}$   & $1/8$ & $0$ & $1/8$ & $0$ & $1/8$ & $0$  & $1/8$  &  $0$ & $1/8$ & $0$  & $1/8$ &  $0$  & $1/8$ & $0$& $1/8$ & $0$  & $1$ &$0$  & $1$ & $0$\\
				$\ket{-}$  &  $\ket{+}$  &  $\ket{-}$  &  $\ket{-}$  & $0$  & $1/8$ & $0$ & $1/8$ & $0$ & $1/8$ & $0$  & $1/8$  &  $0$ & $1/8$ & $0$  & $1/8$ &  $0$  & $1/8$ & $0$& $1/8$   & $1$ & $0$ & $1$ & $1$\\
				$\ket{-}$  &  $\ket{-}$  &  $\ket{+}$  &  $\ket{+}$   & $1/8$ & $0$ & $1/8$ & $0$ & $1/8$ & $0$  & $1/8$  &  $0$ & $1/8$ & $0$  & $1/8$ &  $0$  & $1/8$ & $0$& $1/8$& $0$  & $1$ & $1$ & $0$ & $0$\\
				$\ket{-}$  &  $\ket{-}$  &  $\ket{+}$  &  $\ket{-}$  & $0$  & $1/8$ & $0$ & $1/8$ & $0$ & $1/8$ & $0$  & $1/8$  &  $0$ & $1/8$ & $0$  & $1/8$ &  $0$  & $1/8$ & $0$& $1/8$ & $1$ &  $1$  & $0$ & $1$\\
				$\ket{-}$  &  $\ket{-}$  &  $\ket{-}$  &  $\ket{+}$  & $0$  & $1/8$ & $0$ & $1/8$ & $0$ & $1/8$ & $0$  & $1/8$  &  $0$ & $1/8$ & $0$  & $1/8$ &  $0$  & $1/8$ & $0$& $1/8$ & $1$ &  $1$ & $1$ & $0$ \\
				$\ket{-}$  &  $\ket{-}$  &  $\ket{-}$  &  $\ket{-}$  & $1/8$  &  $0$ & $1/8$ & $0$  & $1/8$ &  $0$  & $1/8$ & $0$ & $1/8$  &  $0$ & $1/8$ & $0$  & $1/8$ &  $0$  & $1/8$ & $0$ & $1$  & $1$ & $1$ & $1$  \\
				\hline
				
				\hline
		        \end{tabular}}
			\label{4_Party_table}
	\end{sidewaystable}
\end{document}